# Do Activists Align with Larger Mutual Funds?*

Manish Jha[†]

November 2024

**Abstract**

This paper demonstrates that hedge funds tend to design their activist campaigns to align with the preferences and ideologies of institutions holding large stakes in the target company. I estimate these preferences by analyzing the institutions' previous proxy voting behavior. The results reveal that activists benefit from this approach. Campaigns with a stronger positive correlation between the preferences of larger institutions and activist communications attract more shareholder attention, receive more votes, and are more likely to succeed.

*I am grateful to Radha Gopalan, Todd Gormley, and Asaf Manela for their continuous guidance and support. I also thank Anat Admati, Taylor Begley, Alon Brav, Stuart Gillan, Jonathan Kalodimos, Mark Leary, Renping Li, Tao Li, James Pinnington, and seminar participants at Georgia State University, Midwest Finance Association, Stanford Rising Scholars Conference, Stevens Institute of Technology, University of North Texas, Virginia Tech, Washington University. Computations were performed using the facilities of the Washington University Center for High Performance Computing.

[†]Georgia State University; 35 Broad Street Office 1241, Atlanta GA 30303; mjha@gsu.edu

> "So the vast majority of companies in the US today are controlled by what I would describe as permanent owners of stock. Think index funds like BlackRock. So the only kind of changes in campaigns we're going to run are ones that benefit the business over decades, and those are the only kind of campaigns you can win. If you have some short-term strategy to make money that's harmful to the company long-term, you're not going to get the support of the BlackRocks, the Vanguards, and the others."
>
> William Ackman, CEO of Pershing Square (NPR 2017)

# 1  Introduction

Although confrontational proxy fights that highlight shareholder preferences occur frequently and attract significant media attention, research on activists' focus on these preferences has been limited. Existing studies primarily concentrate on the characteristics of the parties involved in shareholder activism, their motives, and the interactions between activists and the targeted firms (target). This study aims to shed light on the often-overlooked relationship between activists and shareholders. Specifically, I examine whether activists' proxy communications align with shareholder preferences. The alignment of these preferences is crucial, influencing the issues that activists include in their agendas, as discussed by William Ackman in his NPR interview (shown above). Ex-ante it's unclear how aligning proxy communications with larger shareholders would impact the proxy fight.

Theoretically, catering to larger shareholders allows the activist to consider the preferences of a few institutions, and yet cover a big chunk of the voting bloc. Alchian and Demsetz (1972) specify that the temporary coalescence of share votes into voting blocs is required to displace the existing management or modify managerial policies. Tom Ball, CEO of Vanderbilt Consulting, notes in an interview: "With the increasing concentration of ownership it is the top ten holders who will win it or lose it for you …support on one side or the other of the largest two or three out of the top five will make the difference" (TheStreet 2017). Focusing on a few institutions also reduces the coordination costs of activism (e.g., during the proxy solicitation process).

However, focusing on the larger institutions could also mean a wasted effort. The larger institutions often have a conflict of interest. In particular, a fear of losing the business of corporate pension plans, one of the largest investors in index funds, may deter such institutions from supporting activists ([Ashraf et al. 2012](); [Cvijanović et al. 2016](); [Davis and Kim 2007]()). Additionally, these institutions lack incentives to drive change and are hesitant to expend the resources required for engaging with activists([Lund 2017]()). Despite the SEC ban on selective information disclosure, fund managers might still receive a cold shoulder from executives if they are too critical. Moreover, many larger institutional investors are skeptical of activists' demands, such as requests for increased debt and payouts, which they perceive as shortsighted.

A major hurdle in this line of research is identifying the preferences of institutions. institutions reveal their preferences in many ways, including direct conversations with the target's management and board, exiting positions, and proxy voting. The conversations behind the scenes and the factors that lead institutions to exit positions are not publicly available. As such, researchers have mostly focused on voting record databases to analyze the preferences of institutional investors ([Armstrong et al. 2013](); [Bodnaruk and Rossi 2016](); [Cai et al. 2009]()). These voting records, however, do not have the text content of proposals, which restricts researchers from diving into specifics of the issues on which the voting decisions were made. To overcome this challenge, I extend the voting record database by hand-collecting shareholder proposal texts from the Securities and Exchange Commission (SEC) filings over the 2003–2018 period, which allows me to analyze voting patterns based on proposal content.

I introduce a measure called "Align," which gauges how closely a proxy fight aligns with an institution's preferences on a scale from 0 to 1, based solely on proxy communications. This alignment is determined using a two-step process: (i) I examine the voting decisions of institutions in relation to the textual characteristics of shareholder proposals using a supervised machine learning model known as Support Vector Regression (SVR). The SVR model assigns a coefficient to each phrase based on the institution's past voting patterns. For instance, between 2016 and 2017, DWS consistently voted against management in 97% of climate-related proposals, while BlackRock, Fidelity, and Vanguard opposed management in only 6% of such cases. Consequently, phrases like "climate change" and "environmental concerns" are weighted heavily for DWS in the SVR model. (ii) I combine the frequency and coefficient of each phrase found in the proxy communications to calculate the overall alignment with the institution's preferences. For example, a proxy fight emphasizing climate-related issues and using these specific phrases would be seen as more aligned with DWS's preferences, resulting in an alignment score closer to one.

After measuring the alignment between proxy communications and institutions' preferences, I analyze its correlation with the institutions' ownership of target shares. I find that activists' proxy communications are closely aligned with the preferences of institutions that own larger stakes in the target company. Specifically, a one-standard-deviation increase in target ownership by an institution, which is approximately 0.63 percentage points relative to an average of 0.09%, is associated with about a 0.4- to 0.7-percentage-point increase in proxy communications' alignment with the institution. The average alignment for an institution owning ten percent of target shares is 65%, compared to 46% for an institution with less than 0.01 percent target shares. The increase in support is economically significant and suggests that activists may be writing their communications to solicit support from larger shareholders.

These results suggest that activists cater to the preferences of institutions, but they do not elucidate how these tactics influence proxy fights. Ideally, alignment in proxy fights should benefit activists due to the substantial voting power of these larger institutions. However, as previously discussed, these institutions might simply overlook such tactics due to their limited attention. My findings indicate that campaigns with a stronger positive correlation between the preferences of larger mutual fund institutions and activist communications (i) attract more shareholder attention, (ii) garner more votes, and (iii) have a higher likelihood of success. The following paragraphs will delve into these findings in greater detail.

Theoretical models show that monitors have limited capacity for processing information, so institutions should prioritize proxy fights that align closely with their preferences Kacperczyk et al. (2016); Sims (2003). I find that the institution's attention to a proxy fight, defined as the number of times Internet Protocol (IP) addresses registered under the institution's name access proxy communications filings on the SEC.gov website, increases with the proxy fight's alignment with the institution's preferences. Specifically, a one-standard-deviation increase in the alignment of proxy communications is associated with a 23% increase in the number of times the institution accesses the proxy communications. These results hold even after controlling for the institution's holdings. For similar holdings in targets, the institution is more likely to pay attention to the proxy fight that better aligns with the family's preferences.

Along with higher attention, well-aligned proxy communications are associated with higher support from the institutions. Specifically, among institutions who voted in proxy fight pro-posals, a one-standard-deviation increase in the proxy communications' alignment, is associated with a three percentage points increase in the activist support, relative to an average of 53%. Thus, activists gain institutions' votes when they include fund-specific issues in their communications. The increased support from institutions provides a counterbalance to the ongoing debate about the differing optimization goals of mutual funds and hedge funds. Mutual funds aim for joint portfolio maximization and address agency problems through optimal contract structures (Goetzmann et al. 2003; Lambert and Larcker 2004), investment strategy (Goetzmann et al. 2007), and governance structures (Gillan and Starks 2007; Guercio and Hawkins 1999). In contrast, activists prioritize value maximization and seek to exert direct influence on managerial actions (Edmans et al. 2013; Gantchev 2012), aiming for immediate impacts on share prices (Brav et al. 2015b).

Third, I focus on whether increased attention and support from mutual funds lead to greater success for activists. Brav, Dasgupta, and Mathews (2022) demonstrate in their model that activists can benefit from capital inflows even when their campaigns are not ultimately successful. This suggests that campaigns closely aligned with institution preferences can persist without necessarily leading to higher success rates in proxy fights. My findings indicate that proxy fights aligned with larger shareholders succeed more often. Specifically, among proxy fights with above-average mutual fund ownership, a one-standard-deviation increase in aggregate alignment is associated with a 9.4 percentage point increase in the probability of success for the activist. For reference, activists succeeded in 63% of campaigns from 2004 to 2019.

The proxy fight campaigns that reflect the preferences of larger institutions, as discussed in the paper, could occur through two channels: (i) hedge funds strategically selecting proxy issues or target companies based on anticipated support from larger shareholders, or (ii) institutions anticipating the issues of the campaign, or reaching out to the activists, and increasing their holdings to back the activists. This paper aims to demonstrate that the primary drivers are the former, suggesting that hedge funds tailor their proxy fights to gain more support from significant voting blocs. However, due to the lack of access to the motives and private discussions of hedge fund and mutual fund managers, I cannot definitively rule out the latter. Nevertheless, I attempt to demonstrate that the correlation between alignment and holdings is significant by using pre-proxy fight holdings or by restricting the analysis to a subsample where there is likely an exogenous variation in holdings.

I demonstrate that my results remain robust when using data from six months prior to the proxy fight, indicating that mutual funds' adjustments in holdings do not drive these outcomes in anticipation of the events. Additionally, I examine exogenous variations in institution's holdings in the target company, resulting from mutual fund institution's mergers or changes in the target firm's listing in the Russell 1000 and 2000 indices. On average, I observe a 15% increase in alignment in proxy communications following a merger for institutions that have acquired another institution with investments in the target firm.

Lastly, recognizing that machine learning models can sometimes function as black boxes, I also employ a simpler, manual approach. Specifically, I categorize shareholder proposals into 25 distinct types based on their headings. I then analyze the voting behaviors of institutions within each category over the two years preceding the proxy fight. This method reveals that the positive correlation between institution preferences and activist proposal selection remains consistent even when using a traditional, non-machine learning approach. During their campaigns, activists tend to include proposal types that have historically received favorable votes from major shareholders. I avoid relying on non-machine learning methods for the main findings in this paper because manually categorizing proposals introduces subjectivity. Also, assessing proposals solely based on one-line descriptions overlooks valuable information present in the entirety of each proposal.

Overall, this paper contributes to the literature that studies investor activism. A rich literature studies the characteristics of activists, shareholders, and targets, and the implications of activism for shareholder value and other corporate outcomes.[1] However, relatively little is understood about the interactions between the parties involved, how investors choose their strategies, and what factors contribute to their success. I show that campaign tailoring is an effective way for activists to collectively engage, enabling the small blockholders to govern via voice (Brav et al. 2022). This paper adds to the literature by showing that shareholders' preferences dictate strategies employed by activists, the issues they fight on, the engagement and support they get, and, ultimately, the outcomes of their campaigns.

My research builds on findings by Brav et al. (2023) and Kedia et al. (2021), indicating that activists are more likely to initiate proxy fights in firms with supportive investors. However, my study delves further into the activism process by examining the issues activists raise post-proxy fight initiation. I demonstrate that the language used in these fights aligns with larger institutions and significantly influences the support and success of proxy battles. This complements the work of Appel et al. (2019), who also show that firms' ownership structures play a role in determining the choice of strategies by activists. The granularity of the ownership structure in their paper is restricted to the type of institution, such as active or passive institution. In contrast, I identify preferences at a more granular level based on individual institutions. Institutional investors, even those that belong to the same institution type, have differing ideas of what constitutes the correct course of action.[2]

---

[1]For characteristics of parties involved, see Bradley et al. (2010); Brav et al. (2008); Chapman et al. (2022); Clifford (2008); Greenwood and Schor (2009); Gu and Zhang (2020); Hafeez et al. (2022); Mietzner and Schweizer (2014); Schoenfeld (2020). For the impact of activism, see Boyson et al. (2017); Brav et al. (2015a); Cheng et al. (2012); Ferri and Sandino (2009).

This paper adds to the body of literature investigating voting behaviors within contentious proxy contests. Related works show that investors, who are connected with activists, vote against targets more often (He and Li 2022), and activists regularly interact with asset managers about their plans for a target firm (Edmans and Holderness 2017). Other research finds that mutual funds support target's management when they have business ties (Ashraf et al. 2012; Cvijanović et al. 2016; Davis and Kim 2007), or cross holdings (Harford et al. 2011; Matvos and Ostrovsky 2008). Factors, such as governance failures at mutual funds (Chou et al. 2011) or a common educational background between fund managers and the company's CEO (Butler and Gurun 2012), also add to target-friendly voting. This paper supplements the existing literature by showing that what activists say and the issues they raise also affect shareholders' voting decisions.

Lastly, the paper contributes to a growing body of work that applies text-based analysis to fundamental economic questions (Hanley and Hoberg 2010; Jha et al. 2021, 2022, 2024; Jiang et al. 2019; Loughran and McDonald 2011; Tetlock 2007), in this case quantifying shareholder preference. Prior works employ a more manual approach, classifying proposals into different classes based on issues raised, sponsoring institutions, etc., and subsequently assessing institution voting.[3] Two recent papers that employ statistical techniques to quantify shareholder preference include (i) Bubb and Catan (2021), who undertake a principal components analysis to classify mutual funds in terms of how they follow distinctive philosophies of corporate governance, and (ii) Bolton et al. (2020), who employ a scaling application to place institutions into social-orientation and profit-orientation dimensions. However, the preferences in these papers are based on mutual fund voting patterns with respect to each other and do not take into account the underlying contents of the proposals. This paper is the first to extract proposal's contents and employ a supervised machine learning model to extract mutual fund preferences. The model allows researchers to assess the importance of specific phrases for each mutual fund institution. The model also enables researchers to focus on differences in preferences across institutions and across time.

---

[2]Gao et al. (2021); Matvos and Ostrovsky (2010); Morgan et al. (2011) observe systematic differences in mutual fund voting, indicating divergent preferences. For example, in Trian's proxy fight with Proctor and Gamble, Vanguard sided with the target while BlackRock and State Street voted with the activist, even though they fall in the same type - passive institutions.

[3]Related literature shows that proposals sponsored by institutions get substantially more support, compared to proposals sponsored by individuals (Gillan and Starks 2000), that less myopic funds are more likely to vote for environmental and social issues (He et al. 2023), and that holdings-based corporate social responsibility score for funds is positively associated with voting favorably on social responsibility proposals (Li, Patel, and Ramani Li et al.).

## 2 Data

### 2.1 Shareholder proposal text

I obtain data on institutions' voting from Voting Analytics, which is compiled by Institutional Shareholder Services (ISS). The database includes mutual fund proxy voting records (N-PX filings with SEC). Since July 2003, the SEC has mandated mutual funds to divulge their proxy voting decisions. Thus, the dataset used in this study spans from July 2003 to October 2018. It encompasses both proposals initiated by the firm's management and those put forth by shareholders. To maintain focus and ensure robust analysis, I filter out management-sponsored proposals, as they often entail routine voting and offer less insight into institution preferences (Similar to (Gormley, Jha, and Wang 2024)). By concentrating on shareholder proposals, this approach also addresses the issue of imbalanced datasets, which can arise when one class is disproportionately represented in the training data compared to others.[4]

Next, I add text to the proposals in the voting database. Firms are required to file the Definitive Proxy Statement Form (DEF 14A) with the SEC when a shareholder vote is needed. I match shareholder proposal voting information with DEF14A available via the SEC's Electronic Data

---

[4]Kubat et al. (1997) show that adding examples of majority class could have a detrimental effect on the learner's behavior. For an average proposal (including management sponsored proposals), institutions are unlikely to vote against the management. Over the 2003–2018 period, mutual funds voted against management recommendations for only 9% of management sponsored proposals, as opposed to 42% for shareholder sponsored proposals.

Gathering, Analysis, and Retrieval (EDGAR) system. I use the Central Index Key (CIK) to the Committee on Uniform Securities Identification Procedures (CUSIP) link table, provided by WRDS, to match the two datasets. I supplement this link table with the CIK-CUSIP database, made from parsing 13D/13G filings (Schwartz-Ziv and Volkova 2024). To match the DEF14A proposal text with the voting database, I use: (i) text-similarity of proposal's heading, (ii) proposal sequencing number, and (iii) filing date. In total, I assign 6,176 proposal texts to the shareholder proposals in the voting database. Internet Appendix A.2 explains the process of extracting shareholder proposals and matching them with voting records in detail.

[Table 1 about here.]

Table 1 reports the number of shareholder proposals with matched text for each year from 2003 to 2018. Overall, my sample contains 359 unique institutions, with an average (median) of 1,163 (479) voting observations over the 2003–2018 period. The ISS recommended against the management for 63% of shareholder proposals. As expected, prominent institutions are well diversified and have voted in almost all the shareholder proposals in any particular year. BlackRock voted on 91% of shareholder proposals between 2003 and 2018. Among the top five largest US institutions by assets under management, Vanguard follows management recommendations the most, followed by Fidelity and State Street. The larger institutions, compared to ISS or the overall sample, are less likely to vote against the management. In contrast, smaller institutions, which are often active and follow proxy advisor recommendations, are more willing to show their dissent.

## 2.2 Proxy communications

During a proxy fight, activists and targets put forth their viewpoints to shareholders and send proxy cards to solicit votes. The shareholders sign and return proxy cards to the party they support. Both parties accumulate votes via the returned proxy card and use them at the shareholders meeting. The communications often include a letter to shareholders, which discusses activist's rationale for the proxy fight.

Activists file a Preliminary Proxy Statement in Connection with Contested Solicitations (PREC14A) and a Definitive Proxy Statement in Connection with Contested Solicitations (DEFC14A). Activists also file Additional Definitive Proxy Solicitation Materials Filed by Non-Management (DFAN14A) if the registrant does not support the proxy solicitations. The forms are available to the public via the SEC's EDGAR system. I parse each DFAN, DEFC, and PREC filing (proxy communications filing) to extract the filer and subject company.

I restrict the sample by cross-referencing the filer with a list of investment managers that have filed a Schedule 13F holdings report, a requirement for institutions holding more than $100 million in US stocks at some point in their history. The 13F requirements help exclude activism engagements spearheaded by social activists, disgruntled CEOs, etc.[5]

I do not include 13D filings in proxy communications. In activism literature, 13D filings are often used to identify hedge fund campaigns. These are beneficial ownership filings, required for investors when they own more than 5% in any class of a firm's securities and intend to influence the firm. Although Item 4 of 13D filings has "purpose of the transaction" and sometimes contains activists' intent, activists have incentives to disclose less information, and as such 13D filings often have boilerplate format and do not contain information related to activists' contentions with the manager. Activists often prefer to buy in more than the 5% 13D cutoff and revealing their intentions could increase stock prices. As such, activists tend to not write detailed plans and instead use boilerplate arguments. I also do not rely on media reports to determine the contention, as the sources and linguistic differences add more noise than information.

I start my proxy communications dataset from 2004, six months after the mutual fund voting records are available, to have at least a hundred mutual fund voting records for constructing preferences. I bunch together all the filings for a filer-subject pair if the difference between consecutive filing dates for these documents is less than 180 days. I get a total of 533 confrontational proxy fights over the 2004–2019 period, with an average of 7.9 (median 5) filings per proxy fight. The number of proxy fights involving proxy contests is significantly lower than the general hedge fund campaigns. Activists only use proxy contests as a threat since it is costly for both parties, and only around 10-12% of hedge fund campaigns threaten a proxy contest (Gantchev 2012). In Internet Appendix A.3, I explain the method and textual cues that I employ to extract activists' communication with shareholders.

Table 2 shows notable activists, along with their proxy fights. Some of the well-known ac-tivists such as Icahn Capitals with 39 proxy fights, and Starboard Value with 22 proxy fights, lead the list. Nonetheless, the activism share is fragmented, and only nine activists have double-digit proxy fights over the 2004–2019 period. The 533 proxy fights are shared among 177 unique activists. Often a group of activists together target a firm as a wolfpack (Coffee Jr and Palia 2016; Wong 2019) or coordinate by co-filing Schedule 13Ds (about 22% of Brav et al. (2008) sample). The lead activist files proxy documents with the SEC, and only its name appears in the "filed by" section of the document. Thus, the share of each activist is higher than shown in Table 2.

---

[5] Alexander et al. (2010); Brav et al. (2023); Fos and Tsoutsoura (2014); Norli et al. (2014) employ a similar approach to identify proxy fights. Greenwood and Schor (2009) also use 13F to exclude corporate cross-holdings with activism from portfolio investors.

[Table 2 about here.]

## 2.3 Institution holdings in target

I utilize Center for Research in Security Prices (CRSP) mutual fund holdings data to calculate mutual fund holdings in target stocks. Since 2003, SEC regulations require open-ended mutual funds and ETF portfolios to disclose their holdings quarterly. To aggregate funds at the institution level, I match funds to larger institutions manually, considering subsidiaries within each institution. For instance, Allianz Global Investors acquired Nicholas-Applegate Capital Management and Pacific Investment Management Company in 2000, and invested $2.5 billion in Hartford Financial Services Group in 2008. Thus, I assign funds with names containing "Allianz," "Nicholas-Applegate," "PIMCO," and "Hartford" to the Allianz institution. When aggregating positions at the institution level, I exclude negative fund-level positions, reflecting short positions. Similar findings emerge if I retain these negative positions or use their absolute value for aggregation. In total, I gather holdings information for 438 proxy fights from 2004 to 2019.

To calculate the holdings as a percentage of the target's market capitalization, I utilize the CRSP monthly file. This involves computing each stock's total market capitalization by summing the product of shares outstanding and price across all classes of common stock attributed to a company. For 16 proxy fights where the data is missing in CRSP, I use the book value of common equity from S&P Capital IQ as the market cap. Importantly, the utilization of proxy fight fixed effects ensures that any disparities between book value and market cap do not impact the findings in the paper.

Among the 438 proxy fights analyzed, 46 instances involve an institution possessing more than ten percent of target shares. Additionally, another 142 proxy fights feature at least one institution holding more than five percent of target stocks. On average, mutual funds hold approximately 18.5% of target stocks, with the median holding at 17.6%.

## 3 Method

I exploit institution voting on shareholder proposals to assess investor preferences. This involves analyzing how institutions vote in relation to the textual content of shareholder proposals using SVR. Each phrase in the proposals is assigned a coefficient by the SVR model, derived from the institution's historical voting behaviors. Subsequently, I integrate the frequency and coefficients of these phrases found in proxy communications to determine the extent of alignment with the institution's preferences.

In essence, I train SVR on institutions' support for shareholder proposals to predict their support for proxy communications. However, machine learning models, including SVR, work best when the prediction and the training sample are randomly picked from the same data source. I pick shareholder proposals, as opposed to previous activists' communications, for training because proxy fights are rare. In a typical year, there are less than 20 activist's engagements that lead to a proxy fight. The sparse dataset is not enough to train the machine learning model.

Shareholder proposal texts serve as a viable alternative for proxy communications because they often number in the hundreds each year and cover similar topics. Proxy communications discuss shareholder proposals on which activists are seeking votes. Thus, the selection of shareholder proposals includes those that were voted on during proxy contests. Both types of communication spotlight management's strategic errors and highlight areas for improvement in the firm. Furthermore, over the 2003-2018 period, institutions' voting patterns on shareholder proposals (44% against management) are in line with those on proxy fight proposals (48% against management).

## 3.1 Training SVR on shareholder proposals

To standardize the different ways institutions vote against the management for a proposal, I define a dummy *Align*, which indicates the level of alignment between institution preferences and proposal text. For a particular mutual fund, *Align* is one if the fund does not precisely follow management's recommendation for the proposal. For example, *Align* is one if the management recommendation is "for," and the mutual fund votes "abstain," "do not vote," "withhold," or "against." I use "against management" instead of "for proposal" to be able to extract words that are important enough for the mutual funds to vote against their usual stance of supporting management. The use of "against management" in the training sample also matches well with predicting whether the mutual fund institution will support activists (or vote against management) in the test sample later on. To get institution level *Align*, I average portfolio level *Align* across the institution for the proposal.

I also standardize shareholder proposal texts by removing non-English words, stop words, case, HTML tags, punctuation, digits, inflectional endings, and filler words. I use n-grams of up to five-word phrases to extract features from the text. I omit phrases that appear in less than 1% or more than 70% of the proposals to remove misspelled and frequently used legal terms. I get a total of 9,832 phrases, comprised of 2,465 unigram, 4,991 bigram, 1,593 trigram, 567 4-gram, 216 5-gram. Each shareholder proposal's text is, therefore, represented by $x_s$, a K = 9,832 vector of phrases frequencies, where $x_{p,s}$ = count of phrase p in shareholder proposal text, s. I analyze how the institution voted on the proposal's text, using a linear regression model:

$$Align_{s,i} = \alpha_i + \boldsymbol{\beta_i} \cdot \boldsymbol{x}_s + v_{s,i} \tag{1}$$

where $Align_{s,i}$ is the fraction of funds (managed by institution, i) that voted against the management's recommendation for the shareholder proposal, s. $\boldsymbol{\beta_i}$ is a K vector comprised of a coefficient for each phrase. The high dimensionality of the text makes ordinary least squares and other standard techniques infeasible. To circumvent the problem, I employ a supervised machine learning model - SVR, developed by Drucker et al. (1997). The method is used in accounting and finance literature by Kogan et al. (2009) to predict risk from financial reports, Frankel et al. (2016) to explain accruals, and Manela and Moreira (2017) to measure news implied volatility. The SVR estimation procedure performs well for short samples with a large feature space K. Internet Appendix B.1 describes the process and model parameters I choose in more detail, Internet Appendix D.2 shows that the results in the paper are robust to changing SVR parameters, and Internet Appendix B.3.1 shows an example to support that SVR coefficients are based on institutions' past voting outcomes.

The cost of SVR is that it cannot concentrate on sub-spaces of **x** (Hastie et al. 2009). For example, if an institution is environmentally conscious and votes against management when specific phrases such as "climate change" occurs in proposal text, the SVR will assign a high positive coefficient to "paris." Even though the phrase "paris" is orthogonal to voting decisions in most cases, it gets a positive coefficient from co-occurrence with "climate change" in "Paris Agreement on Climate Change." Overall, SVR performs well in predicting institution voting behavior. The effectiveness of a machine learning algorithm is determined by its ability to predict out-of-sample observations. The out-of-sample mean absolute error for 'Align' for an average institution is 0.40. By using coefficients in the proposal text, SVR reduces the mean absolute error to 0.24.

## 3.2 Measuring communications' alignment with institution preferences

To determine institution preferences for a particular proxy fight, I analyze its voting choices on the shareholder proposals from the first proxy communication filing date (proxy fight date) to two years before the proxy fight date. To have sufficient training samples, I only consider institutions that have voting information for at least a hundred proposals during the period. For example, in the P&G 2017 proxy fight by Trian Funds, the first proxy (between DEFC, DFAN, and PREC) filed by Trian was a DFAN14A on July 17th, 2017. So, I consider shareholder proposals during the July 17th, 2015–July 17th, 2017 period. A total of 189 institutions voted in at least a hundred of these shareholder proposals.

For each institution, I solve SVR regression Equation 2, which assigns a coefficient to each of the phrases in shareholder proposals. Using frequencies of phrases in the proxy communication and their associated coefficients, I estimate the proxy communication's, p, alignment with institution, i, preferences or $Align_{p,i}$ as:

$$Align_{p,i} = \alpha_f + \boldsymbol{\beta}_f \cdot \boldsymbol{x}_p \tag{2}$$

where $\boldsymbol{\beta}_f$ is the estimated K-vector containing coefficients assigned to phrases derived from training the SVR model on the institution's voting history. The proxy communication is rep-resented by **x**, a K-vector of phrases' frequencies in the proxy communication. The estimated alignment, or $Align$, is the likelihood of an institution supporting the activist based solely on proxy communication and is bounded by zero and one. Thus, if an activist uses phrases such as "climate change," which are important to DWS, indicated by the phrase's positive coefficient in DWS's SVR model, the proxy communication's alignment with DWS preferences will be closer to one.

## 3.3 Sample and descriptive statistics

For my primary analysis, I use the proxy communications' alignment with each institution who has voted in at least a hundred shareholder proposals in the two years before the proxy fight date. Some of these institutions are not invested in the target; in which case, I assign an equity share of zero. My sample includes 522 proxy fights (438 with at least one invested institution), involving 287 unique institutions, over the 2004–2019 period. In total, my sample contains 66,836 observations (12,582 with non-zero holdings).

[Table 3 about here.]

Table 3 provides an overview of the sample set, consisting of 66,432 observations from proxy fights by institutions. On average, institutions hold 0.09% of the target's market cap, with a standard deviation of 0.63%. Only 18.9% of the sample shows a non-zero holding, and among these cases, institutions hold 0.49% of the target's market cap. Regarding the outcome variable *Align*, the mean value is 0.48, indicating that, on average, proxy communications align with institution preferences by 48%. This means the activist can expect support from 48 out of 100 mutual fund portfolios within an institution during the campaign. The standard deviation for *Align* is 0.4. Specifically, 19.2% of the sample has *Align* equal to zero, while 21.0% have *Align* equal to one. The text's alignment bunching at zero and one is in line with how most portfolios in an institution vote as a block (Cai and Walkling 2011; Rothberg and Lilien 2006).

I focus on the dataset containing all the institutions. The full dataset takes care of the survivorship bias in the subsample of institutions that hold shares in the target. I hypothesize that activists care more about the preferences of institutions that own more shares. For instance, an activist would align their proxy communications more with an institution holding a 5% stake than one with a 0.1% stake. However, what if the proxy communications align more with institutions that don't have any investment? To mitigate concerns about survivorship bias, I concentrate on the entire dataset. Separately, Internet Appendix C.2 reports results for the subsample of institutions with a stake in the target. The appendix also shows that the results hold if we exclude observations where the institution owns more than 5% of the targeted firm.

[Figure 1 about here.]

Figure 1 plots the average estimated proxy communications' alignment text for different institution holdings. The bulk of the dataset is in the left corner, with only 278 (or 2.19%) observations having an institution that owns more than 5% of target shares. Across holdings, the average alignment is above 40%. The plot has an increasing trend, indicating a positive association between institutions owning shares in the target and proxy communications catering to institutions' preferences. For the subsample of BlackRock and Vanguard, the increasing trend in the proxy communications' alignment when holdings increase persists. The SVR method predicts, on average, BlackRock, compared to Vanguard, is more likely to support activists. BlackRock's higher activist support is in line with how it had voted more aggressively against management recommendations in shareholder proposals, illustrated in Table 1. The figure suggests that activists include issues important to the institutions that hold more shares.

## 4   Results

To examine the relationship between the alignment of activist communications and institution holdings, I estimate the following equation:

$$Align_{p,i} = \beta Holding_{p,i} + \delta_p + \delta_i + \epsilon_{p,i} \tag{3}$$

where $Align_{p,i}$ is the predicted alignment of the proxy communications, *p*, with the institution's, *i*, preferences. *Holding* refers to the fraction of the target's market cap the institution owns before a proxy fight. $\delta_p$ and $\delta_i$ are proxy fight level and institution level fixed effects. Because both *Align* and *Holding* could be correlated across observations of a particular proxy fight and because the estimation errors, $\epsilon$, might exhibit serial correlation, I cluster the standard errors at the proxy fight level. However, subsequent findings are robust to not clustering or clustering at other levels (e.g., institution).

The primary concern in this research design is the issue of omitted variables. If the institution's holding in target is correlated with activist-, target-, or institution-level characteristics that affect proxy communications' alignment with the family, then my estimate of interest, $\beta$ could be biased. This bias would arise because $\beta$ might capture the influence of these omitted variables rather than the true effect of holdings on campaign tailoring. For instance, if activists tend to align their proxy fights with major institutions based on total net assets (e.g., BlackRock, Vanguard), and these institutions typically hold significant shares in an average proxy fight, then the variables *Holding* and *Align* could be positively correlated. This correlation could occur not because activists intentionally align their proxy fights with larger shareholders, but because the largest institutions inherently possess significant holdings in these companies.

The inclusion of proxy fight and institution fixed effects allows me to control for several of these potential omitted factors. Proxy fight fixed effects account for activists' characteristics that could influence the tailoring of proxy communications to align with institution preferences, such as activists' skills and ISS support. They also control for target characteristics (e.g., performance, strategy, ownership structure) at the time of the proxy fight that might affect institutional reactions to specific phrases. Institution fixed effects control for differences in an institution's overall tendency to vote against management, which can vary significantly across institutions (Brav et al. 2023; Kedia et al. 2021). Thus, by including both fixed effects, the coefficient of interest, $\beta$, is identified using the variation in how proxy communications' alignment for a given proxy fight varies with each institution's holdings and how an institution's alignment varies with its holdings in targeted firms' stocks.

[Table 4 about here.]

Table 4 reports estimates for the association of the predicted proxy communications' alignment on institution holdings using variants of Equation 3. The percentage of shares held by institutions in the target is significantly (at 1% level) associated with the proxy communications' alignment with the family. A one-standard-deviation increase in institution holdings in targeted shares, which is approximately 0.63 percentage points relative to an average of 0.09%, is associated with around a 0.47 percentage point increase in the proxy communications' alignment with the institution preferences. Relative to the average alignment of 48%, this corresponds to a sizable increase. The coefficient for holdings is positive and significant at either level of fixed effects. Separately, Internet Appendix Table 14 reports similar results for sub-sample datasets (i) including only institutions invested in the target and (ii) excluding observations where an institution owns more than five percent of the market cap of the targeted firm.

An alternative scenario that could explain the positive association - if larger institutions, such as BlackRock, Fidelity, and Vanguard, generally have a higher alignment with activist proxy communications. However, Section 2.1 suggests otherwise. Institutions with higher holdings, often passive index investors, are less likely to vote against management in shareholder proposals. Since the training data is not skewed, this pattern is unlikely to appear in the predicted or test samples. Table 4 further supports this hypothesis with regression results that include the institution fixed effect. Specifically, for a particular institution, the proxy communications' alignment increases by 1.1 percentage points (1.59×0.724) for every percent increase (1%/0.63% = 1.59 standard deviation) in the family's holdings in the target. This suggests that the text is tailored towards the preferences of major institutional holders in the target, rather than major institutional investors in general. The Internet Appendix C.1 provides examples of activists using specific phrases to appeal to BlackRock, Fidelity, and Vanguard when these institutions hold larger stakes in the target.

## 5 Affects of proxy communication alignment

### 5.1 Alignment is positively associated with institution's attention

I use institutions' proxy communications filings accessed on the SEC's EDGAR server as an indicator of their attention to proxy fights. The SEC's Division of Economic and Risk Analysis (DERA) compiles data on internet search traffic for EDGAR filings through SEC.gov, covering the period from February 14, 2003, to June 30, 2017. I use a linking table from Digital Element to assign IP addresses in the log files to institutions. The linking table contains organizations' names and registered IP addresses as of December 31st, 2016. I follow Iliev et al. (2021) to identify EDGAR activity related to governance research. Using the accession numbers in the proxy communications filings, I compile a list of proxy fight documents. I then total the number of times an institution viewed any of these proxy fight documents from the start date of the proxy fight to 30 days after its conclusion. The start and end dates of the proxy fight are defined by the first and last filing dates of the proxy communications.

In total, I gather data on fund views of proxy communications for 427 proxy fights, involving 115 unique institutions. For each proxy fight, I include institutions that meet the following criteria: (i) they have voted in at least one hundred shareholder proposals in the two years prior to the proxy fight, and (ii) they have checked filings for at least 1% of their investments. My data set comprises 244,000 proxy fight-institution pairs, with an average of 1.04 views by an institution per proxy fight. Focusing only on positive views, I have 40,000 proxy fight-institution data points, involving 73 unique institutions across 278 proxy fights. The average number of positive views per institution is 6.39 views per proxy fight. The Internet Appendix A.4 provides a detailed description of the procedure used to extract proxy communications filings accessed by the institutions on the SEC server.

[Figure 2 about here.]

Figure 2 illustrates the differences in attention among institutions based on the alignment of proxy communications. When a proxy fight is more aligned, indicated by an alignment score above 0.5, institutions are more likely to access proxy communication filings on SEC.gov. This pattern holds consistently across various institutions' holdings in the target company. For comparable investments, institutions are more inclined to pay attention when proxy communications address their concerns. To formally test whether higher alignment is associated with increased attention, I estimate:

$$View_{p,i} = \beta Align_{p,i} + \delta_p + \delta_i + \epsilon_{p,i} \tag{4}$$

where $View_{p,f}$ represents the number of times proxy communications, $p$, were accessed by an institution, $i$. $Align_{p,i}$ is the predicted alignment of proxy communications with the institution preferences. $\delta_p$ and $\delta_i$ are proxy fight level and institution level fixed effects. Finally, I adjust the standard errors, $\epsilon_{p,i}$, for clustering at the proxy fight level. The model is estimated over the January 2004–June 2017 period.

[Table 5 about here.]

Table 5 illustrates that institutions pay more attention to proxy fights that align with their preferences. The estimated coefficient is positive and statistically significant at the 1% level, regardless of whether proxy fight-level or institution-level variations are considered. Specifically, a one-standard-deviation increase in the predicted alignment of proxy communications with institution preferences—approximately 40 percentage points relative to an average alignment of 48%—is associated with an increase of 0.11 views of proxy communication filings on the SEC.gov website. Given that the average number of views for a proxy fight is 0.48, this increase represents a 23% rise in EDGAR activity by the institution. This heightened attention has significant implications for the proxy fight, as institutions typically exhibit a nonchalant attitude toward activism. For instance, BlackRock, Fidelity, and Vanguard chose to forgo voting in the GameStop ballots by keeping their shares on loan (WSJ 2020b).

Drake et al. (2020); Gormley and Jha (2024); Iliev et al. (2021); Malenko and Shen (2016) demonstrate that institutions pay more attention when they have a higher stake in a firm. Moreover, Section 4 shows that the alignment is dependent on the equity share the fund owns. To mitigate the omitted variable problem, I include the institutions' equity share in the target as a control in part (4), (5), and (6) of Table 5. Even for similar investments in the target, institutions to which proxy communications is more aligned pay more attention to proxy filings. Internet Appendix Table 15 reports the result of the analysis of the smaller sub-sample of invested institutions.

## 5.2 Alignment is positively associated with institution's support

To evaluate the impact of proxy communications on institutions' voting behaviors, I focus my analysis on proxy battles that progressed to the voting phase. Utilizing data from CapitalIQ provided by S&P, I compile information on the outcomes of these proxy fights. I get results for 461 proxy fights within my sample. The CapitalIQ platform classifies proxy fight results into four categories: (i) Successful, where the activist's proposals secure victory in the shareholder election; (ii) Settled, indicating negotiations and compromises between targets and activists without a formal election, often prompted by the target's recognition of the activist's strong case; (iii) Withdrawn, denoting instances where activists perceive inadequate support and opt to withdraw the case; and (iv) Unsuccessful, wherein activists participate in the election but fail to garner the necessary votes.

[Figure 3 about here.]

Figure 3 shows the distribution of proxy fight outcomes. Across the period from 2004 to 2019, 63% of these battles culminate in either successful resolutions or settlements, with both categories peaking at 85% for fights originating in 2009. This proportion remains consistently above 60% throughout the analyzed period, with the exception of proxy battles commencing in 2011. Notably, only approximately half of the proxy fights proceed to proxy voting. Gantchev (2012) estimates the average cost of a confrontational proxy fight to be $10.71 million. Additionally, targets are concerned about reputation costs, particularly in the event of election losses, motivating both activists and targets to seek settlements outside of proxy contests. Over the 2004-2019 period, 199 proxy fights proceeded to actual shareholder voting, resulting in outcomes classified as either Successful or Unsuccessful. This figure aligns closely with the 207 voted proxy fights documented in Brav et al. (2023). Utilizing the voting data from these proxy fights, I compare text-based voting predictions with the actual voting decisions of institutions.

To identify the proposals associated with a proxy fight, I start with all the shareholder meetings for the target. I apply filters to exclude meetings without shareholder-sponsored proposals, those occurring more than 30 days before the end date of the proxy fight, or those happening more than 365 days after the proxy fight's end date. The 30-day window is chosen because activists often communicate with shareholders post-voting to relay meeting outcomes and express gratitude. Following filtration, I select the first meeting after the proxy fight's initiation date. From this selection, I extract all shareholder proposals lacking the phrase "Management Nominee" in the ISS voting database. I then aggregate activist support at the proxy fight level; for instance, if an institution backs one out of three activist proposals, the *SupAct* for the proxy fight is recorded as 0.33. This process yields a total of 1,457 voting records detailing institution participation in proxy fight proposals. To evaluate my hypothesis that institutions exhibit a more favorable voting pattern toward proxy fights aligning with their preferences, I conduct the following estimation:

$$SupAct_{p,i} = \beta Align_{p,i} + \delta_p + \delta_i + \epsilon_{p,i} \tag{5}$$

where $SupAct_{p,i}$ is the fraction of mutual funds, for an institution, *i*, that supported activists' proposals in the shareholder meeting following a proxy fight, *p*. $Align_{p,i}$ is the proxy com-munications' alignment with institution preferences. $\delta_p$ and $\delta_i$ are proxy fight level and institution level fixed effects. Finally, I adjust the standard errors, $\epsilon_{p,i}$, for clustering at the proxy fight level.

[Table 6 about here.]

The results, presented in Table 6, demonstrate a positive association between the alignment of proxy communications and institution preferences, and the actual voting outcomes. The estimated coefficients are positive and statistically significant (at the 1% level) for proxy fight and institution level fixed effect. Among the institutions participating in proxy fights, a one-standard-deviation increase in alignment between the proxy fight and institution preferences—equivalent to a 44 percentage point deviation from the average 56% alignment—is correlated with a three percentage point rise in the institution's actual support for the activist, relative to the average of 53%. Hence, the text-based measure of voting outcomes consistently mirrors the actual voting trends, remaining robust even when considering variations within proxy fights or fund levels. These findings bolster the arguments put forth by Kamenica and Gentzkow (2011); McCloskey and Klamer (1995), who argue that persuasion plays a key role in voting decisions. A higher movement required for the text's alignment to change the institution's actual support indicates that other factors, apart from proxy communications content, dictate institution voting. These factors, along with the limitations inherent in a text-based measure, are further discussed in Internet Appendix C.4.

## 5.3 Well-aligned proxy fights are more likely to succeed

Securing support from a significant shareholder lends credence to a campaign, often swaying other shareholders to rally behind it. A notable example occurred in 2019 when T. Rowe Price, EQT's largest shareholder, issued a press release stating support for dissident Rice Group nominees. Subsequently, at EQT's annual meeting, all seven Rice-nominated directors were elected by shareholders. This raises the inquiry: does alignment of proxy communications with major investors increase the likelihood of success? To address this question, I commence with a straightforward measure of how effectively an activist aligns proxy communications with the preferences of institutions. I define:

$$AgAlign_p = \sum_f Align_{p,i} \times \frac{Holding_{p,i}}{\sum_f Holding_{p,i}} \tag{6}$$

where *AgAlign* is aggregate align for a proxy fight, *p*, based on the proxy fight's alignment with an institution weighted by the institution's holdings. It measures aggregate mutual fund support, i.e., what fraction of the mutual fund's vote will the activist gather, based on the proxy communications. A proxy fight that is well-aligned with institutions owning larger shares in the target will have a higher *AgAlign*.

A drawback of *AgAlign*, as a parameter for proxy fight outcome, is that it is volatile for proxy fights where the mutual fund presence is not significant. For a proxy fight where mutual funds own less than five percent of the market cap, even when the proxy fight is well-aligned with *AgAlign* close to one, the small ownership is usually not sufficient to have a proxy fight level impact. Brav et al. (2022) also show in their model that campaigns succeed if the measure of shares that engage is above a threshold. To circumvent the issue, I interact *AgAlign* with a dummy for ownership in the target. The ownership dummy, *OwnDum*, is one if the mutual funds, whose alignments aggregate to *AgAlign*, own more than the sample average, 14.3%, in target shares. Out of the 419 proxy fights that had an institution with voting records as a shareholder, 195 have ownership dummy of one. Internet Appendix Figure 11 shows that the result holds for changing the cutoff of ownership dummy.

[Figure 4 about here.]

Figure 4 is a scatter plot for aggregate alignment, averaged each year based on the proxy fight's outcome. I divide the proxy fight outcomes, shown in Figure 3, into two groups: Settled/Successful and Withdrawn/Unsuccessful, as the outcomes within the groups are considered equivalent in activism literature (Brav et al. 2015b). Proxy fights that more closely align with the preferences of larger institutions tend to succeed more frequently. Each year, except 2015, proxy fights that end in success exhibit a higher average aggregate alignment. Within a sub-sample of twelve proxy fights where an activist engaged the same target twice with differing outcomes, the average aggregate alignment for Successful/Settled proxy fights stands at 56%, compared to 39% for Unsuccessful/Withdrawn proxy fights. To formally assess whether the alignment of proxy communications correlates with activist success, I employ:

$$Win_p = \gamma AgAlign_p + \lambda OwnDum_p + \beta AgAlign_p \times OwnDum_p + \epsilon_p \tag{7}$$

where $Win_p$ is a dummy equal to one if the activist wins the proxy fight (Successful or Settled). *AgAlign* is the aggregate mutual fund support for the activist, based on proxy communications. *OwnDum* denotes the ownership dummy, which is one if the mutual funds, whose alignment aggregate to *AgAlign*, own more than the sample average, 14.3%, in target shares. By including the additional interaction, the coefficient on *AgAlign* will now capture the importance of text alignment for all the other proxy fights. In contrast, the sum of the coefficients on *AgAlign*, *OwnDum*, *AgAlign* × *OwnDum* will capture the importance of text alignment for the proxy fights with above-average mutual fund holdings. The standard errors, $\epsilon_p$, are robust and computed with the sandwich estimator of variance.

[Table 7 about here.]

Table 7 shows that proxy fights, which are more in line with larger institutions' preferences and have sufficient mutual fund holdings, are indeed more likely to succeed, as shown in Column (2). Specifically, among proxy fights for which the mutual fund ownership is at least the average, a one-standard-deviation increase in *AgAlign*, or 28 percentage points, is associated with a 9.4 (11.1 + 0.48 - 2.18) percentage point increase in the likelihood of the activist winning the proxy fight. For reference, the average *AgAlign* is 53% for the sample, i.e., for an average proxy communications, the activist could expect 53% of all the votes cast by mutual funds. The 9.4% increase is significant given the average *Win*, or a proxy fight's likelihood of being successful or settled is 63%. The coefficient for ownership dummy is not significant, indicating that proxy fight outcomes are not significantly different across the *OwnDum* cutoff. The relationship also does not hold for proxy fights that do not have mutual fund investments above the 14.3% threshold, illustrated in Column (1).

# 6 Robustness to Alternative Explanations

## 6.1 Using pre-proxy fight holdings pattern

Campaigns that reflect the preferences of larger institutions can occur through (i) hedge funds selecting targets or proxy issues that align with those of larger shareholders, or (ii) funds anticipating campaign issues and increasing their holdings to support the activists. In Table 4, I argue in favor of the former.

[Table 8 about here.]

To test my hypothesis, I re-run the regressions from Table 4 using institution holdings from six months earlier. Table 8 shows that the alignment of proxy communications with institution preferences remains significant even when considering these six-month lagged holdings. These results hold because institutions typically do not make significant changes to their portfolios before proxy fights.

## 6.2 Restricting samples to exogenous holdings shocks

In this section, I focus on a subset of activism events where institution holdings in the target company may have changed due to exogenous factors during the activism period.

### 6.2.1 Mergers between mutual fund institutions

I begin with a list of mutual fund institutions' mergers from Lewellen and Lowry (2021) and supplement this with mergers reported in the media. My focus is on proxy fights that (i) occurred during periods when activists were sharing proxy communications, and (ii) involved target companies in which the acquired mutual fund institution held shares.

I identify 11 such proxy fights and split each into two observations: one for the period before the merger and one for the period after. I assign proxy communications to the "before" or "after" category based on their timing relative to the merger. Subsequently, I measure *Align* for all institutions involved in each proxy fight before and after the merger. I create two dummy variables: (i) *Acquired*, which is set to 1 if an institution acquired another institution with investments in the target firm during the proxy fight period, and (ii) *Post*, which is set to 1 for proxy fights that include proxy communications occurring after the merger event.

[Table 9 about here.]

Table 9, shows that, within this subsample, the Acquired-Post interaction is positively associated with *Align* after controlling for proxy fight and institution fixed effects. On average, proxy communications are 15% more aligned post-merger for institutions that acquired another institution with investments in the target firm. I also implement a staggered difference-in-difference setup in accordance with Baker et al. (2022), appending the dataset with addi-

tional proxy fights that did not involve a merger event in Column 2. The estimates remain consistent.

### 6.2.2 Russell index reconstitution

Similarly, for Russell index reconstitution, I search for proxy fights where the target firm's stock assignments changed between the Russell 1000 and 2000 indexes during the activism period. Since 2007, Russell Investments has adopted a "banding" policy, making it more challenging for firms to switch between the two indexes ([Appel et al. 2020](#)). Due to the limited number of observations, I do not observe a significant difference in institution ownership before and after the Russell index switch overall. Instead, I conduct a case study focusing on cases where a target firm gained more than 1% of market cap investments from institutions after switching to the Russell 2000 index.

I find one such proxy fight involving Pentwater Capital (activist) and Leap Wireless (target), where Leap Wireless was reassigned from the Russell 1000 index to the Russell 2000 index in 2011. I identify three such cases: Allianz, BlackRock, and Calamos Investments. I split proxy communications that were filed before (or after) the Russell index reconstitution date, and measure *Align* for each of the institution. Figure [5](#) illustrates that institutions increased their share of holdings in the target after its switch to the Russell 2000 index. Concurrently, the alignment of proxy communications from the activist also increased for these institutions after the Russell reconstitution.

[Figure 5 about here.]

## 6.3 Using a non-machine learning method

While the SVR method is interpretable in the sense that one could examine the coefficients of each phrase and it's based on the institution's voting choices, it can be challenging to track. In my analysis, the SVR involves coefficients for more than nine thousand phrases, which determine the alignment of proxy communications. As a result, it becomes cumbersome to keep track of all the moving parts. In this section, I aim to verify whether the positive association between a proxy communication's alignment and institution ownership persists when using a simpler non-machine learning method to measure alignment.

Although the ISS voting database does not include the full text of proposals, it does provide a one-line description, typically the heading, for each proposal. I categorize the most common descriptions into 25 proposal types, such as director election, governance, sustainability, etc. By starting with the most common proposal types, I am able to classify 90% of the shareholder proposals into one of these 25 categories. Internet Appendix D.1 lists the classification of proposal descriptions into different types. I use the method described in Section 5.2 to get proposals related to a proxy fight.

For this analysis, I define alignment slightly differently. Firstly, I calculate an institution's alignment with a proxy fight proposal in two steps: (i) I identify shareholder proposals of the same type as the proxy fight proposal with meeting dates within two years prior to the beginning date of the proxy fight.(ii) I determine the fraction of these shareholder proposals in which the institution voted against management recommendation. To assess the proxy fight's alignment with the institution preferences, I then average the alignment of proxy fight proposals within each proxy fight.

[Table 10 about here.]

Table 10 presents the results of the re-evaluation of Section 4 using a proposal-type-based measurement of alignment. A one-standard-deviation increase in institution holdings of targeted shares, approximately 0.66 percentage points relative to an average of 0.1%, is associated with an increase of around 0.6 percentage points in the alignment of proxy communications with institution preferences. These coefficients are comparable to those in Table 4. The findings suggest that activists strategically focus on proposal types where larger shareholders have voted against management.

I refrain from utilizing the non-machine learning method as the primary setup throughout the paper due to the inherent subjectivity associated with manually classifying proposals into types. Moreover, analyzing proposals based solely on their one-line descriptions excludes additional information that could be gleaned from the entirety of the proposal. For instance, a proposal with the headline "director election" might contain details in the body such as "the candidate has extensive experience with climate-related risks." In such cases, while the manual method may only register the institution's support for directors, a machine learning model would discern that the institution supports proposals that advance climate-related issues.

# 7 Conclusion

In financial markets, institutions overseeing mutual funds play a crucial role, yet they often face constraints, both legal and incentive-based, when interacting with portfolio managers. Critics argue that these institutions tend to be overly deferential towards portfolio managers, potentially neglecting the interests of the fund's investors (Bebchuk and Hirst 2019). Lund (2017) goes as far as to suggest that index mutual funds should refrain from voting altogether, leaving such decisions to more actively engaged parties.

This study delves into the impact of shareholder preferences on activist campaigns, strategies, and their outcomes. In their proxy communications, I find that activists' use of language is positively associated with the preferences of institutions that own a larger share of the targeted firm. Using institution voting histories as a measure of preference, I find a positive correlation between alignment with these preferences and the extent of fund ownership. For every percentage increase in fund holdings, there is a corresponding 0.7 percentage point increase in alignment with institution preferences. This implies that activists may strategically raise issues that resonate with larger shareholders.

Proxy communications that are well-aligned with an institution's preferences are associated with higher institution's attention and votes. A one-standard-deviation increase in proxy communications alignment is associated with a 23% increase in institution attention and a 6% increase in actual activist support. These proxy fights are also more likely to end up in favor of the activists.

The finding raises concerns about the disproportionate influence a few shareholders can wield over corporate decision-making. WSJ (2020a) notes that "markets are shifting from harnessing the wisdom of crowds to the wisdom of a handful of influential money-management executives." As a prevention mechanism against anti-competitive influence, the SEC has different filing requirements based on the investors' desire to influence. While an institution looking for change has to file the stringent beneficial ownership form 13D and disgorge profit on trades for six months (Section 16B), an institution in the ordinary course of business needs to file a much less stringent form 13G (Morley 2018). However, this study suggests that activists serve as conduits through which fund preferences can shape corporate governance.

While I validate my findings using proxy guidelines texts and alternative methodologies, I lack counterfactual data to explore different scenarios. What if activists employed alter-native communication proxies? How might that have impacted outcomes? Moreover, how would external shocks, such as changes in target ownership during a proxy fight, influence activist discourse? Unfortunately, the current dataset on activism lacks the depth to delve into these questions. Additionally, this study focuses solely on one dimension of persuasion within activism, neglecting other important factors like private shareholder-target meetings or activists' media interactions. Future research should examine how external shocks and varied communication modes shape persuasion dynamics in shareholder activism.

Wong, Y. T. F. (2019, November). Wolves at the Door: A Closer Look at Hedge Fund Activism. *Management Science 66*(6), 2347–2371.

WSJ (2018). BlackRock, Calpers Want Exchanges to Clamp Down on Dual-Class Shares. [https://www.wsj.com/articles/blackrock-and-calpers-to-stock-exchanges-clamp-down-on-dual-class-shares-1540394503](https://www.wsj.com/articles/blackrock-and-calpers-to-stock-exchanges-clamp-down-on-dual-class-shares-1540394503).

WSJ (2020a). Replacing the wisdom of crowds with the wisdom of fink. [https://www.wsj.com/articles/replacing-the-wisdom-of-crowds-with-the-wisdom-of-fink-11579429800](https://www.wsj.com/articles/replacing-the-wisdom-of-crowds-with-the-wisdom-of-fink-11579429800).

WSJ (2020b). WSJ News Exclusive | How Investing Giants Gave Away Voting Power Ahead of a Shareholder Fight. [https://www.wsj.com/articles/how-investing-giants-gave-away-voting-power-ahead-of-a-shareholder-fight-11591793863](https://www.wsj.com/articles/how-investing-giants-gave-away-voting-power-ahead-of-a-shareholder-fight-11591793863).
34

**Figure 1:**
**Alignment is positively associated with holdings**
The figure plots the average proxy communications' alignment with institution preferences for holdings between 0 to 10%. The alignment is based on the institution's voting patterns on shareholder proposals in the two years before the proxy fight. Holding represents the institution's ownership in the target stock as a percentage of the target's market cap. Holdings are rounded to the nearest tick mark, and the corresponding alignments are averaged. The radius for all institution series corresponds to the number of observations around the tick. To flesh out numbers close to zero, I plot the x-axis in logarithmic terms.

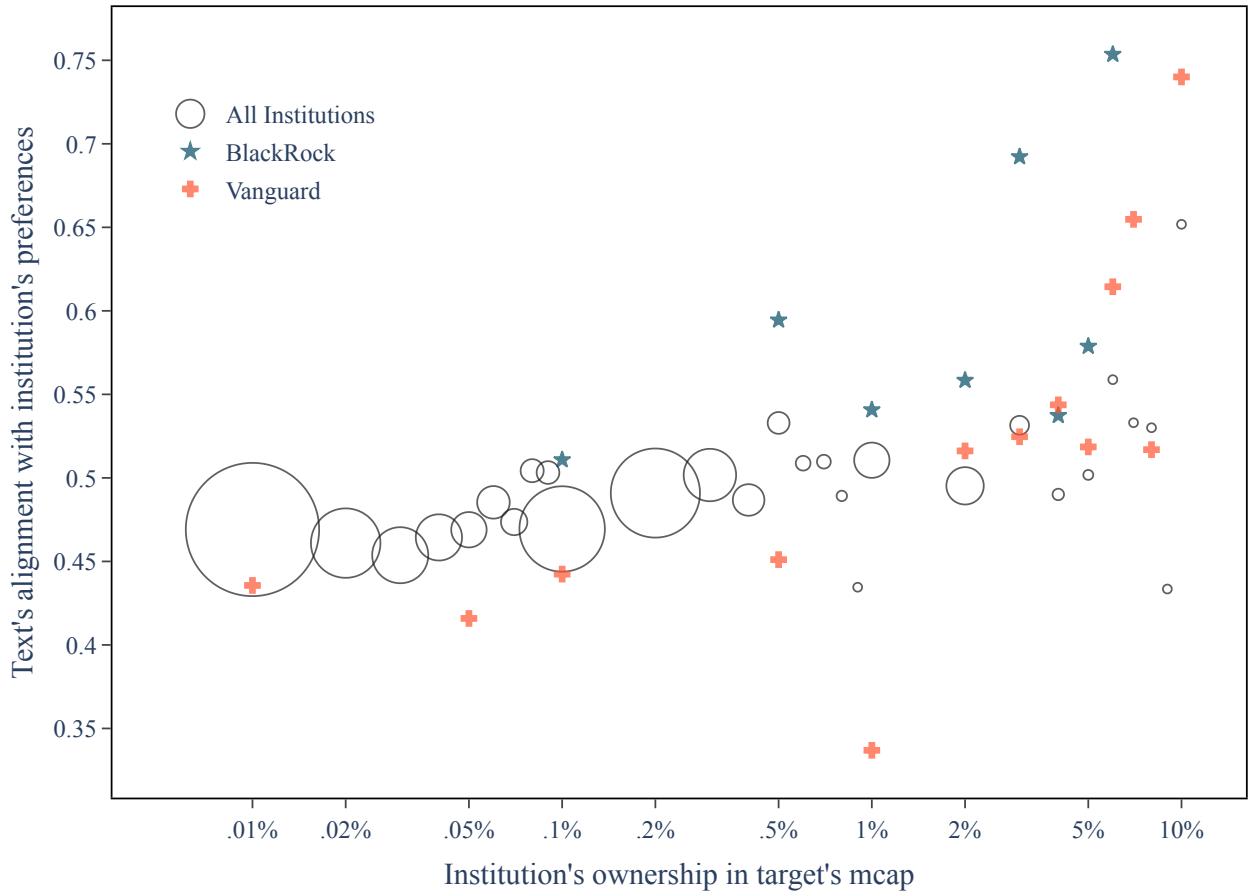

**Figure 2:**
**Institutions conduct more research about proxy fights that are well-aligned.**
This figure plots the number of times institutions accessed proxy communications filings on the SEC.gov server, averaged at each half percentage point holdings. The data for institutions' access of SEC filings is available from DERA. The period considered for each proxy fight spans the date the proxy fight begins to 30 days after the proxy fight ends. The proxy fight's beginning (end) date is the first (last) date of proxy communications filing by the activist. Proxy communications' alignment with institution preferences is based on the family's voting choices on shareholder proposals.

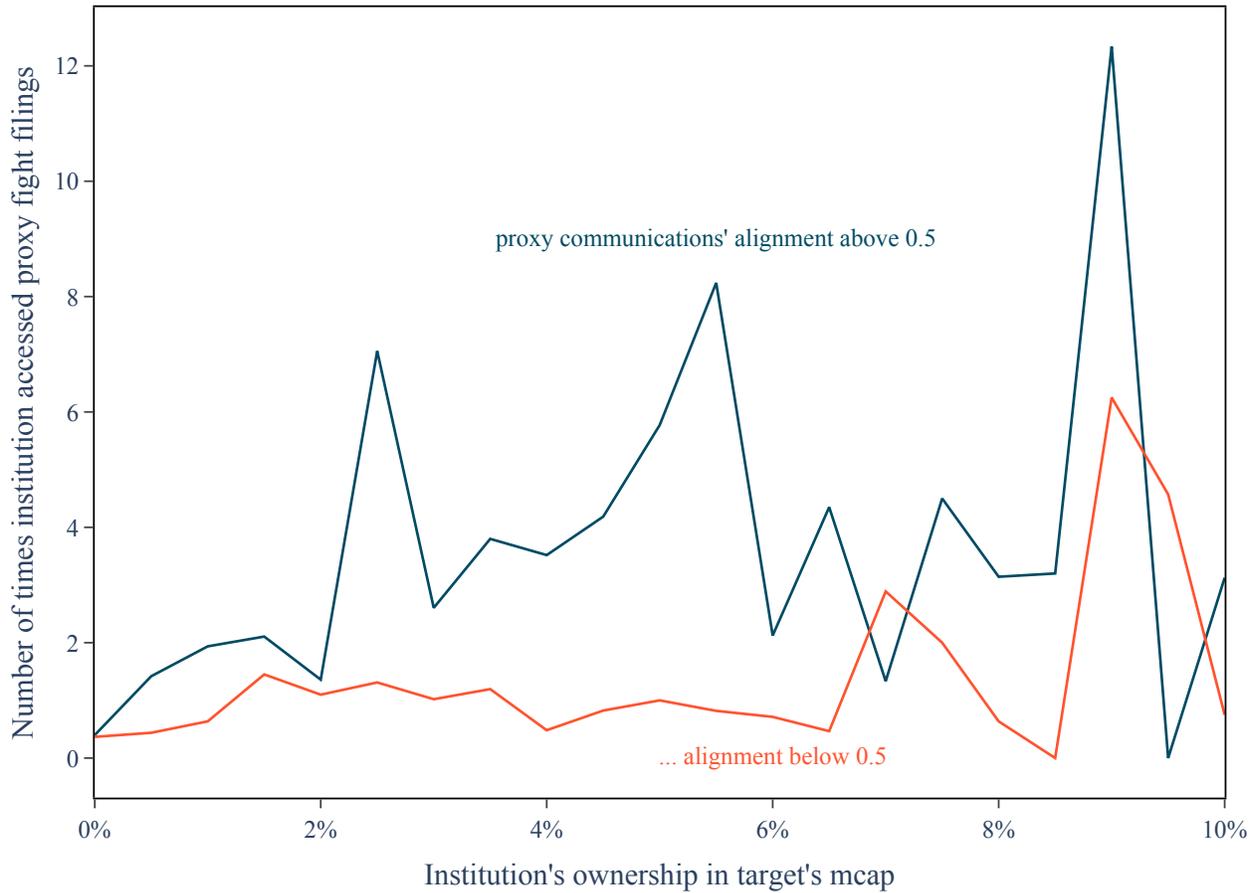

**Figure 3:**
**Distribution of proxy fight outcomes remains persistent.**
The stacked area plots the outcome of proxy fights over the 2004–2019 period. Proxy fights are assigned to the year when they began, i.e., the earliest date of SEC filings pertaining to the proxy fights. The information on proxy fights' outcomes is collected from S&P CapitalIQ.

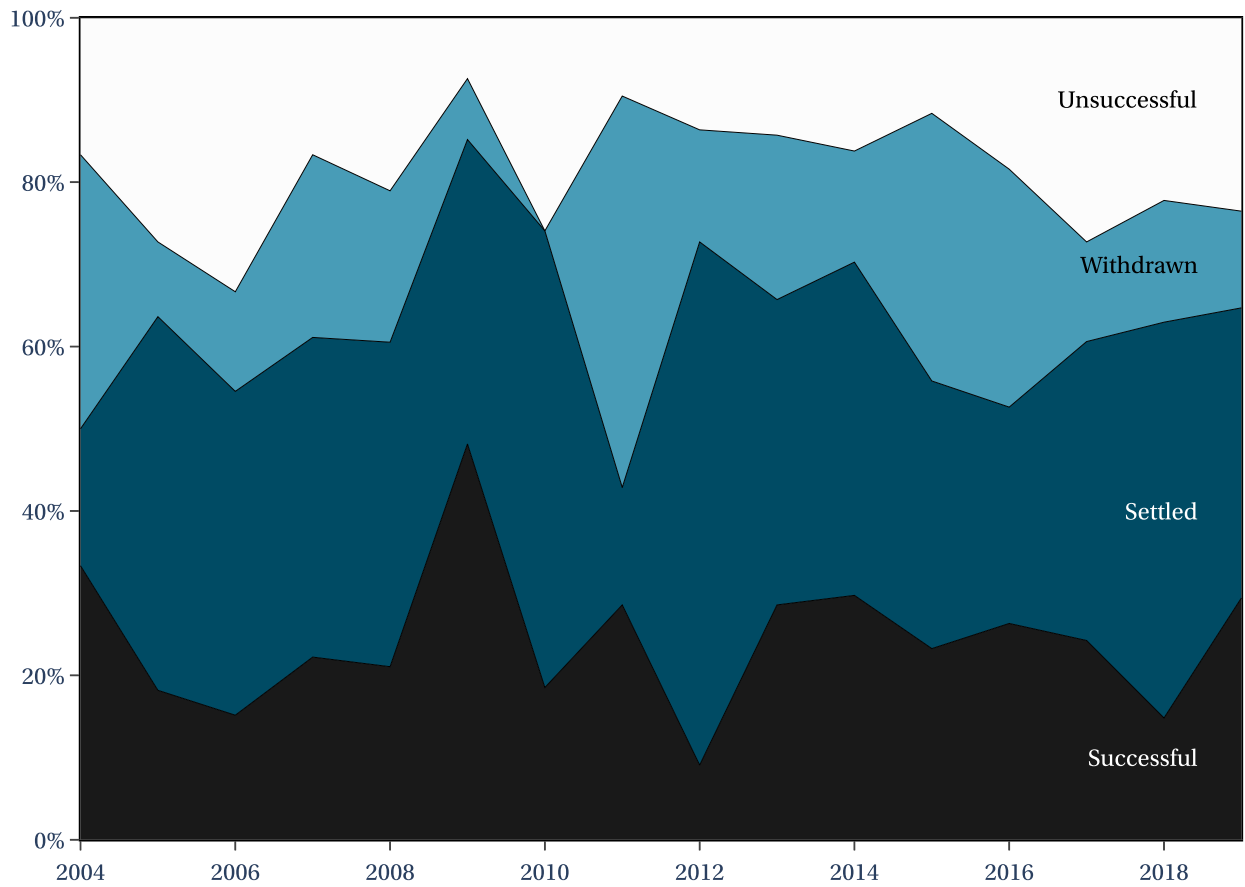

**Figure 4:**
**Proxy fights that are aligned well with the larger shareholders are more likely to win.**
This figure plots the proxy communications' aggregate alignment, averaged for each year based on the proxy fight's outcome. The sample includes proxy fights with at least 14.3% (the mean holding) of target shares held by one of the institutions with available voting records. The aggregate alignment is the proxy communications' alignment weighted by institution holdings, defined in Equation 6. The shade of the bubble represents the outcome of the proxy fight. In 2006 and 2010, no proxy fights above the cutoff holding were unsuccessful or withdrawn.

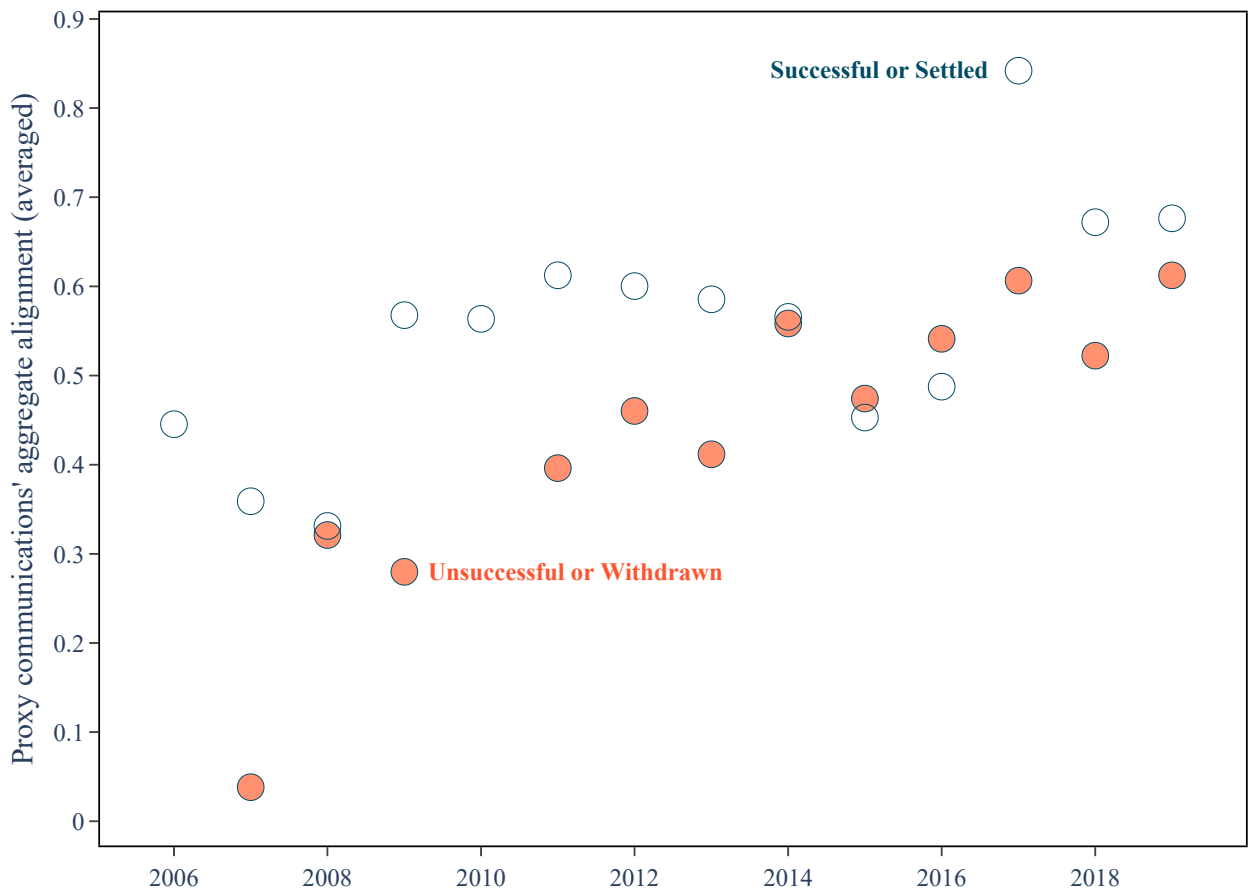

**Figure 5:**
**Alignment follows increased investments after Russell reconstitution**
The figure plots holdings and alignment for the proxy fight between Pentwater Capital (activist) and Leap Wireless (target) in 2011, before and after Leap Wireless assignment into the Russell 2000 index from the Russell 1000 index.

**(a)** Institution holdings in Leap Water, before and after Russell reconstitution.

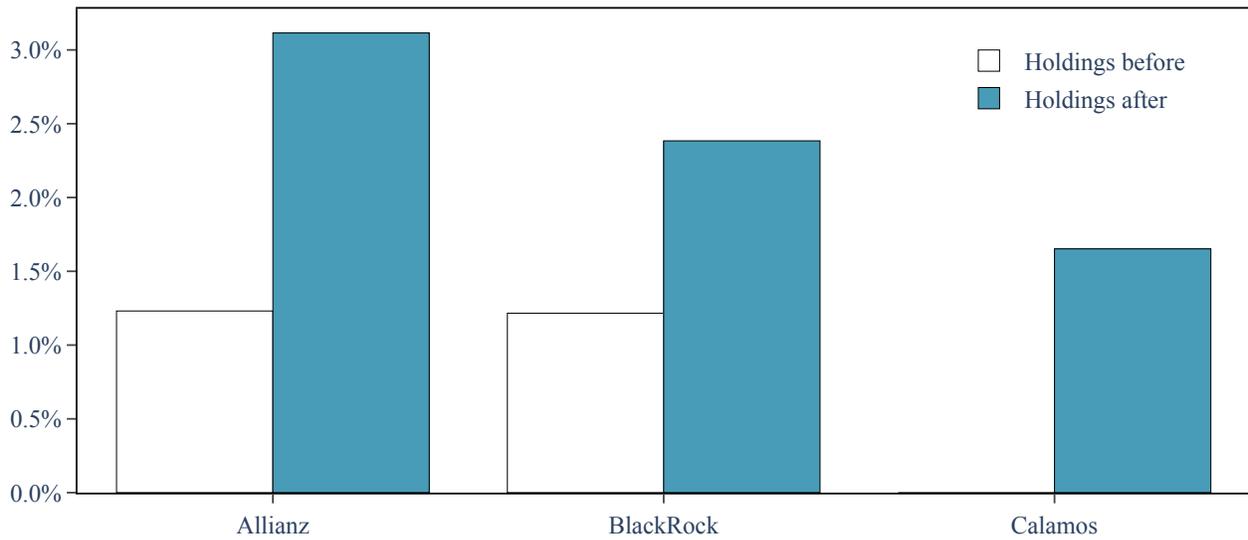

**(b)** Alignment of Pentwater Capital's proxy communications, before and after Russell reconstitution.

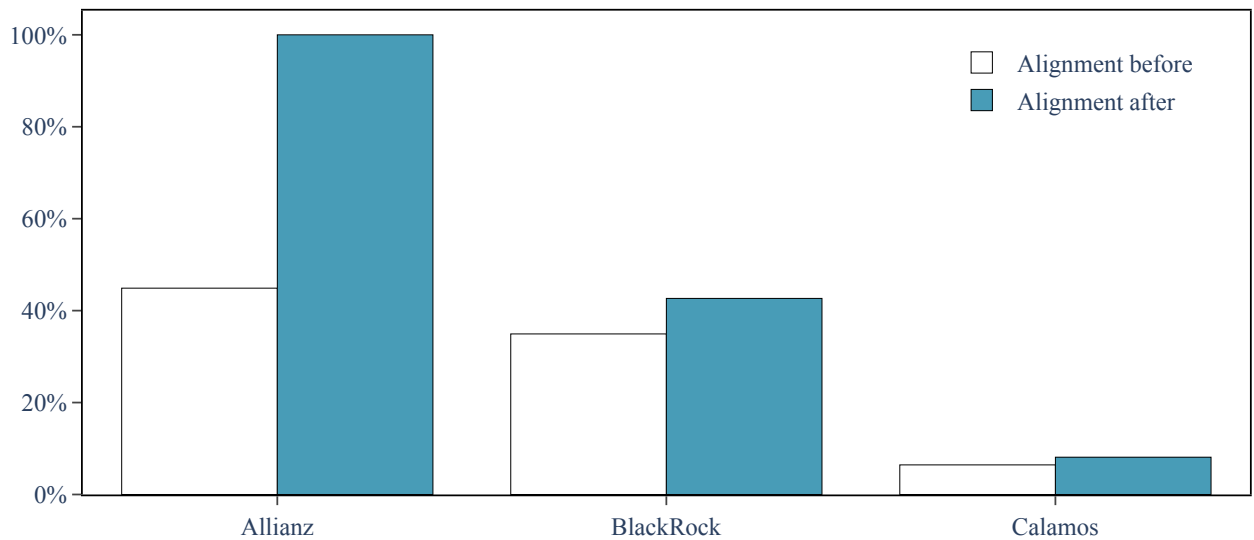

**Table 1:**
**The largest institutions tend to follow management recommendations**
This table reports institution voting on shareholder proposals by year. The sample contains all the shareholder proposals for which the text was available from the SEC. In Column (1), the number in parentheses indicates the percent of proposals for which ISS recommended against the management. Columns (2), (3), (4), (5), and (6) show the voting history of the five largest US institutions by asset under management (RelBanks 2017). The number inside parentheses indicates the percent of proposals with an against management vote during the year.

| Year | Shareholder Proposals (1) | BlackRock (2) | Charles Schwab (3) | Fidelity (4) | State Street (5) | Vanguard (6) | Full Sample (7) |
|---|---|---|---|---|---|---|---|
| 2003 | 42 (40) | 24 (27) | 15 (20) | - | - | - | 841 (29) |
| 2004 | 499 (45) | 439 (27) | 343 (16) | 15 (42) | 7 (14) | 17 (34) | 15,692 (31) |
| 2005 | 480 (50) | 452 (35) | 380 (16) | 400 (26) | 337 (10) | 425 (40) | 24,777 (31) |
| 2006 | 488 (62) | 440 (39) | 483 (57) | 479 (24) | 83 (16) | 483 (19) | 37,437 (42) |
| 2007 | 398 (60) | 364 (47) | 384 (62) | 344 (27) | 46 (11) | 380 (18) | 27,674 (41) |
| 2008 | 374 (62) | 261 (52) | 330 (65) | 334 (37) | 324 (21) | 313 (26) | 21,228 (45) |
| 2009 | 529 (72) | 406 (56) | 510 (68) | 523 (40) | 494 (30) | 459 (16) | 39,973 (52) |
| 2010 | 360 (77) | 344 (33) | 301 (49) | 327 (36) | 317 (22) | 327 (20) | 24,980 (51) |
| 2011 | 257 (78) | 249 (45) | 193 (63) | 226 (46) | 221 (36) | 226 (34) | 14,598 (59) |
| 2012 | 358 (65) | 328 (41) | 242 (49) | 329 (39) | 304 (42) | 292 (31) | 21,577 (51) |
| 2013 | 497 (65) | 492 (27) | 412 (37) | 494 (23) | 464 (35) | 478 (18) | 34,142 (44) |
| 2014 | 534 (62) | 530 (20) | 418 (30) | 521 (22) | 516 (39) | 521 (17) | 38,383 (41) |
| 2015 | 573 (71) | 538 (31) | 476 (17) | 540 (22) | 533 (38) | 533 (15) | 51,438 (44) |
| 2016 | 413 (66) | 391 (24) | 360 (19) | 389 (19) | 380 (37) | 393 (17) | 34,507 (41) |
| 2017 | 371 (60) | 353 (27) | 296 (25) | 346 (28) | 331 (30) | 355 (20) | 30,497 (40) |
| 2018 | 3 (100) | 2 (0) | - | 2 (0) | 2 (0) | 3 (0) | 104 (41) |
| Total | 6,176 (63) | 5,613 (35) | 5,143 (40) | 5,269 (29) | 4,359 (31) | 5,205 (21) | 417,848 (44) |

**Table 2:**
**Only a few activists have led more than ten proxy fights**
This table reports prominent lead activists and firms they targeted over the 2004–2019 period. Some of the larger target firms, in terms of their market cap, are also listed.

| Activists (# proxy fight) | Major Targets |
| --- | --- |
| Breeden Cap Mgmt (2) | Applebees, PIMCO |
| Icahn Enterprises (39) | Time Warner, Yahoo, Dell, eBay, AIG, Clorox, Family Dollar, Motorola, Tyson, Xerox, Cigna, Biogen |
| Land & Buildings Inv Mgmt (8) | Macerich, MGM Resorts, Taubman Centers |
| Marcato Cap Mgmt (3) | Lear Corp, Deckers Outdoor Corp, Buffalo Wild Wings |
| P Schoenfeld Asset Mgmt. (3) | T-Mobile |
| Pershing Square Cap Mgmt (2) | Allergan Inc, Automatic Data Processing |
| Starboard Value (22) | Bristol Myers, Office Depot, Dollar Tree, Yahoo, AOL |
| Steel Partners (11) | Rowan Companies, GenCorp |
| Third Point (9) | Dow Chemicals, Campbell, Yahoo |
| Trian Funds (3) | P&G, DuPont, Heinz |

**Table 3:**
**Summary statistics for proxy fight-by-institution sample**
This table reports summary statistics for proxy fight-by-institution level outcome and explanatory variable. Holding is the fraction of the target's equity owned by an institution, and $Align$ is the predicted proxy communications' alignment with institution preferences. To match later estimations, the sample is limited to institutions that have voted in at least a hundred shareholder proposals in the two years before a proxy fight.

|         | Mean   | Median | SD     | % of observations with non-zero value | Mean if non-zero | Number |
|---------|--------|--------|--------|----------------------------------------|------------------|--------|
| Holding | 0.0009 | 0      | 0.0063 | 18.9 %                                 | 0.0049           | 66,432 |
| $Align$ | 0.48   | 0.43   | 0.40   | 80.8 %                                 | 0.59             | 66,432 |

**Table 4:**
**Activists communications are positively correlated with preferences of larger shareholders**
This table reports estimates of regression of the proxy communications' alignment with institution preferences on the institution's holdings in targets. Specifically, I estimate:

$$Align_{p,i} = \beta Holding_{p,i} + \delta_p + \delta_i + \epsilon_{p,i}$$

where $Align_{p,i}$ is the predicted alignment of the proxy communications, $p$, with the institution, $i$, preferences. *Holding* is the percent of equity the institution owns of the target before the proxy fight. $\delta_p$ and $\delta_i$ represent proxy fight level and institution level fixed effects, respectively. The sample consists of proxy fights, identified using proxy communications filings, over the 2004–2019 period. Corresponding institutions include all the institutions that have voted in at least a hundred shareholder proposals in the two years prior to the proxy fight. The independent variable is scaled by the standard deviation of the underlying variable, meaning the coefficient can be interpreted as the effect of a one-standard-deviation change in the determinant. Standard errors, $\epsilon_{p,i}$, are clustered at the proxy fight level, and *t*-statistics are reported in brackets below the coefficient estimates. The symbols *, **, and *** indicate significance at 10%, 5%, and 1% respectively.

|  | Proxy communications' alignment with institution preferences | | | |
|---|---|---|---|---|
|  | (1) | (2) | (3) | (4) |
| Fraction of target mcap held by institution | 0.0059*** [3.81] | 0.0042** [2.19] | 0.0073*** [3.15] | 0.0047** [2.48] |
| Proxy fight FE |  | Yes |  | Yes |
| Institution FE |  |  | Yes | Yes |
| Observation | 66,432 | 66,432 | 66,432 | 66,432 |
| $R^2$ | 0 | 0.135 | 0.094 | 0.224 |

**Table 5:**
**Institutions conduct more research on proxy fights that are close their preferences**
This table reports estimates of regression of institution access of proxy communications filings on the proxy communications' alignment with institution preferences. Specifically, I estimate:

$$View_{p,i} = \beta Align_{p,i} + \delta_p + \delta_i + \epsilon_{p,i}$$

where $View_{p,f}$ is the number of times an institution, $i$, accessed proxy communications filings, $p$, between the date the proxy fight begins to 30 days after the proxy fight ends. The proxy fight's beginning (end) date is based on the first (last) date of proxy fight filing by the activist. $Align_{p,i}$ is the proxy communications' alignment with institution preferences. $\delta_p$ and $\delta_i$ are proxy fight level and are institution level fixed effects. Data for institutions' access of filings on SEC.gov is available via DERA. Columns (4), (5), and (6) control for institution holdings in the target. Independent variables are scaled by the standard deviation of the underlying variable, meaning coefficients can be interpreted as the effects of a one-standard-deviation change in the determinant. Standard errors, $\epsilon_{p,i}$, are clustered at the proxy fight level, and $t$-statistics are reported in brackets below the coefficient estimates. The symbols *, **, and *** indicate significance at 10%, 5%, and 1% respectively.

|  | Number of times institution viewed proxy communications filings on SEC.gov | | | | | |
| --- | --- | --- | --- | --- | --- | --- |
|  | (1) | (2) | (3) | (4) | (5) | (6) |
| Proxy communications' alignment | 0.0856*** | 0.0849*** | 0.115*** | 0.0774*** | 0.0771*** | 0.112*** |
|  | [4.24] | [3.59] | [4.69] | [3.86] | [3.36] | [4.57] |
| Fraction of target mcap held by institution |  |  |  | 0.381*** | 0.365*** | 0.251*** |
|  |  |  |  | [18.98] | [7.01] | [5.25] |
| Proxy fight FE |  | Yes | Yes |  | Yes | Yes |
| Institution FE |  |  | Yes |  |  | Yes |
| Observation | 34,174 | 34,174 | 34,173 | 34,174 | 34,174 | 34,173 |
| $R^2$ | 0.001 | 0.106 | 0.163 | 0.011 | 0.115 | 0.167 |

**Table 6:**
**Institutions support is positively associated with proxy communications alignment**
This table reports estimates of regression of institution activist support on the proxy communications' alignment with institution preferences. Specifically, I estimate:

$$SupAct_{p,i} = \beta Align_{p,i} + \delta_p + \delta_i + \epsilon_{p,i}$$

where $SupAct_{p,i}$ is the fraction of mutual funds, for an institution, $i$, that supported activists' proposals in the shareholder meeting following a proxy fight, $p$. Proxy fight proposals include shareholder proposals that were part of shareholder meeting after the proxy fight. $Align_{p,i}$ is the proxy communications' alignment with institution preferences. $\delta_p$ and $\delta_i$ are proxy fight level and institution fixed effects. The sample contains observations of institutions voting on proxy fight proposals over the 2004–2019 period. The independent variable is scaled by the standard deviation of the underlying variable, meaning the coefficient can be interpreted as the effect of a one-standard-deviation change in the determinant. Standard errors, $\epsilon_{p,i}$, are clustered at the proxy fight level, and $t$-statistics are reported in brackets below the coefficient estimates. The symbols *, **, and *** indicate significance at 10%, 5%, and 1% respectively.

|  | Actual activist support | | |
| --- | --- | --- | --- |
|  | (1) | (2) | (3) |
| Proxy communications' alignment | 0.0186* [1.88] | 0.0468*** [4.44] | 0.0310*** [3.2] |
| Proxy fight FE |  | Yes | Yes |
| Institution FE |  |  | Yes |
| Observation | 1457 | 1453 | 1419 |
| $R^2$ | 0.002 | 0.557 | 0.611 |

**Table 7:**
**Activists are more likely to succeed when the communication is well aligned**
This table reports estimates of regression of proxy fight outcomes on proxy communications' alignment weighted by the institution holdings. Specifically, I estimate:

$$Win_p = \gamma AgAlign_p + \lambda OwnDum_p + \beta AgAlign_p \times OwnDum_p + \epsilon_p$$

where $Win_p$ represents a dummy, which is one if the result of the proxy fight, $p$, is Successful or Settled. *AgAlign* is the holdings-weighted proxy communications' alignment with institutions. *OwnDum* is the ownership dummy, which is one if the mutual funds with voting information own more than the average, 14.3%, of target shares. The sample consists of all the proxy fights, identified using SEC filings, that went to a voting stage over the 2004–2019 period. The independent variable, *AgAlign*, is scaled by the standard deviation of the underlying variable, meaning the coefficient can be interpreted as the effect of a one-standard-deviation change in the determinant. The standard errors, $\epsilon_p$, is robust and computed with the sandwich estimator of variance. The symbols *, **, and *** indicate significance at 10%, 5%, and 1% respectively.

|  | Indicator for activist win | |
|---|---|---|
|  | (1) | (2) |
| Aggregate alignment | 0.0214 | −0.0218 |
|  | [0.90] | [-0.71] |
| Ownership dummy |  | 0.0048 |
|  |  | [0.10] |
| Aggregate alignment × |  | 0.111** |
| Ownership dummy |  | [2.25] |
| Observation | 419 | 419 |
| $R^2$ | 0.002 | 0.014 |

**Table 8:**
**The positive correlations remain on using six months prior holdings**

This table reports estimates of regression of the proxy communications' alignment with institution preferences on the institution's six month lagged holdings in targets. Specifically, I estimate:

$$Align_{p,i} = \beta Holding6m_{p,i} + \delta_p + \delta_i + \epsilon_{p,i}$$

where $Align_{p,i}$ is the predicted alignment of the proxy communications, $p$, with the institution, $i$, preferences. *Holding6m* is the percent of equity the institution owns of the target six months before the proxy fight. $\delta_p$ and $\delta_i$ represent proxy fight level and institution level fixed effects, respectively. The sample consists of proxy fights, identified using proxy communications filings, over the 2004–2019 period. Corresponding institutions include all the institutions that have voted in at least a hundred shareholder proposals in the two years prior to the proxy fight. The independent variable is scaled by the standard deviation of the underlying variable, meaning the coefficient can be interpreted as the effect of a one-standard-deviation change in the determinant. Standard errors, $\epsilon_{p,i}$, are clustered at the proxy fight level, and $t$-statistics are reported in brackets below the coefficient estimates. The symbols *, **, and *** indicate significance at 10%, 5%, and 1% respectively.

|  | Proxy communications' alignment with institution preferences | | | |
|---|---|---|---|---|
|  | (1) | (2) | (3) | (4) |
| Target mcap held by institution, | 0.0022 | 0.0020*** | 0.0022*** | 0.0019*** |
| 6-months prior to proxy fight | [1.42] | [12.04] | [3.80] | [7.43] |
|  |  |  |  |  |
| Proxy fight FE |  | Yes |  | Yes |
| Institution FE |  |  | Yes | Yes |
|  |  |  |  |  |
| Observation | 66,432 | 66,432 | 66,432 | 66,432 |
| $R^2$ | 0 | 0.134 | 0.094 | 0.224 |

**Table 9:**
**Proxy communications get more aligned to institutions after acquisitions**
This table reports estimates of regression of the proxy communications' alignment with institution preferences on the institution's acquisition holdings in targets. Specifically, I estimate:

$$Align_{p,i} = \beta AcquiredxPost_{p,i} + \delta_p + \delta_i + \epsilon_{p,i}$$

where $Align_{p,i}$ is the predicted alignment of the proxy communications, $p$, with the institution, $i$, preferences. Acquired is a dummy variable that turns 1 if an institution acquired another institution with investments in target during the activism period, and Post is a dummy that equals 1 for proxy fights during which a merger occurred and contains proxy communications after the merger event. $\delta_p$ and $\delta_i$ represent proxy fight level and institution level fixed effects, respectively. For Column 1, the sample consists of proxy fights, identified using proxy communications filings, over the 2004–2019 period during which a merger between institutions occurred. For Column 2, the sample is expanded appending all the other proxy fights during which no merger between institutions occurred. Corresponding institutions include all the institutions that have voted in at least a hundred shareholder proposals in the two years prior to the proxy fight. Standard errors, $\epsilon_{p,i}$, are clustered at the proxy fight level, and $t$-statistics are reported in brackets below the coefficient estimates. The symbols *, **, and *** indicate significance at 10%, 5%, and 1% respectively.

|  | Proxy communications' alignment with institution preferences | |
|---|---|---|
|  | (1) | (2) |
| Acquired x Post | 0.1574* | 0.1508*** |
|  | [1.79] | [2.58] |
|  |  |  |
| Proxy fight FE | Yes | Yes |
| Institution FE | Yes | Yes |
|  |  |  |
| Includes non-merger proxy fights |  | Yes |
|  |  |  |
| Observation | 3882 | 68,662 |
| $R^2$ | 0.352 | 0.226 |

**Table 10:**
**Activists include proposal types that are well-aligned with larger shareholders**
This table reports estimates of regression of a proxy communications' alignment with institution preferences on the institutions' holdings in targets. Specifically, I estimate:

$$Align_{p,i} = \beta Holding_{p,i} + \delta_p + \delta_i + \epsilon_{p,i}$$

where $Align_{p,i}$ is the predicted alignment of proxy communications, $p$, with the institution, $i$, preferences. *Align* is calculated based on the family's voting in shareholder proposals that of the same type as proxy fight proposals. *Holding* is the percent of equity the institution owns of the target before the proxy fight. $\delta_p$ and $\delta_i$ represent proxy fight level and institution level fixed effects, respectively. The sample consists of proxy fights, identified using proxy communications filings, that went to a voting stage over the 2004–2019 period. Corresponding institutions include all the institutions that have voted in at least a hundred shareholder proposals in the two years prior to the proxy fight. The independent variable is scaled by the standard deviation of the underlying variable, meaning the coefficient can be interpreted as the effect of a one-standard-deviation change in the determinant. Standard errors, $\epsilon_{p,i}$, are clustered at the proxy fight level, and *t*-statistics are reported in brackets below the coefficient estimates. The symbols *, **, and *** indicate significance at 10%, 5%, and 1% respectively.

|  | Proxy communications' alignment with institution preferences (simpler method) | | | |
| --- | --- | --- | --- | --- |
|  | (1) | (2) | (3) | (4) |
| Fraction of target mcap | 0.0270*** | 0.0288** | 0.0038 | 0.0064* |
| held by institution | [10.67] | [9.04] | [1.16] | [1.97] |
| Proxy fight FE |  | Yes |  | Yes |
| Institution FE |  |  | Yes | Yes |
| Observation | 13,328 | 13,328 | 13,326 | 13,326 |
| $R^2$ | 0.008 | 0.136 | 0.370 | 0.494 |

# Internet Appendix

## A  Data

### A.1  Largest institutions invested in targets

Figure 6a illustrates the list of the largest institutions invested in targets at the initiation of a proxy fight. Vanguard is the largest shareholder in 155 proxy fights, followed by Fidelity at 58, and BlackRock at 52. The list of largest shareholders contains fifty unique mutual fund institutions. It seems like the usual suspects such as BlackRock, Fidelity, and Vanguard are the major shareholders in all the proxy fights; and activists have to simply align communications with them irrespective of the proxy fight. However, Figure 6b shows otherwise. The top three institutions are the largest shareholders in 61% of the proxy fights. The distribution of holdings for the top three institutions demonstrates that these institutions do not play a significant part in many of the proxy fights. BlackRock, Fidelity, and Vanguard own less than 1% target share in 47%, 65%, and 36% proxy fights, respectively. The distribution underlines that the institutes holding voting power vary across targets. As such, the activists have to tailor their approach for each proxy fight, instead of catering to the same few institutions across proxy fights.

[Figure 6 about here.]

### A.2  Assigning proposal's text to ISS voting data

Voting records of the institution, at the mutual fund level, are available from the ISS. I aggregate mutual fund voting information into institution voting data, based on the names of the mutual funds, mergers and acquisitions, and investment relationships among mutual fund institutions. During the period 2003–2018, I have 359 institutions who have voted in at least a hundred proposals. These institutions voted on a total of 10,679 unique shareholder proposals. For the text of the proposals, I use DEF14A filings, which are available in the EDGAR system via the SEC. The system provides indexes to all public filings, including CIK, type of form, filing date, and weblink.

    I match proposals in the ISS voting database to the text available in the DEF14A filings. The voting data provides a record date, the meeting date, proposal item number, and a short

description of the proposal. To make a suitable match, I start by slicing the shareholder proposal for a particular CIK and subsequently for a specific meeting date. To get a list of potential text matches for a proposal in voting data, I employ a two-step process. First, I gather all the proposals for a particular CIK on a meeting date in the voting data. Next, I slice the SEC index file for the particular CIK, DEF14A filing, and filing date between the record and meeting date. The average number of proposals for a firm on a meeting date is 2.17. Usually, proposals about director elections are grouped as one in DEF14A filings. Therefore, for searching in DEF14A, I combine all the director election proposals into a single proposal.

I parse the DEF14A HTML file using [Beautiful Soup](#) python package. I remove all the tables, white space, accented characters, and non-UTF encoding. I also filter out the first 75, which has filer information, and the last 75 lines, which are often errors from PDF to HTML conversion, from the filings. Once I have the clean DEF14A text, I look for sections of the filing that correspond to the specific proposal. To get the starting line for a proposal, I assign a score to each line of the DEF14A based on how likely it matches the ISS proposal description and item number. I choose the line with the maximum score. I assign higher scores if the line (i) is uppercase, (ii) contains words such as proposal, number, no., item, etc. (iii) contains the same words as it appears in the ISS description (iv) has less than 80 characters (v) contains the same number as ISS item number. Sometimes the proposals are written in two lines - the first line containing the item number and the second containing the description. To take this into account, I repeat the same process by combining two consecutive lines and checking the score improvement.

To find the starting line for the next proposal, I begin five lines after the previous proposal's start line. Proposals in DEF14A are typically sequentially put; thus, I choose the ending for a proposal as two lines before the start of the next one. To get the last proposal's ending line, I begin five lines after the start and look for the phrase "The Board of Directors recommends." If there are no matches, I take the ending line as fifty lines after the starting line. I assign the text between the starting line and the ending line in DEF14A to each proposal. For director election proposals, which generally have one proposal for all the nominated directors, I choose paragraphs between the starting line and the ending line that contains the name of the director listed in the voting database.

Out of the 10,679 shareholder proposals, I assign text to 6,176 proposals. The difference in numbers is because (i) the ISS data includes shareholder proposals for companies across the globe, while SEC filings are done by US-based companies (ii) some of the proposals are written in a nonstandard format, which makes parsing them precisely difficult (iii) given the goal of this study is to analyze the text, I limit the sample to proposals, where I am able to match text information with reasonable confidence, and which has more than 30 words.

## A.3  Extracting text associated with a proxy fight

To get the proxy communication, I look for forms DEFC14A, DFAN14A, and PREC14A filed by the investment firm and parse filer and subject company CIKs. Two fields characterize the proxy filings associated with proxy fights: (i)*FILED BY*, containing the activist information, and (ii)*SUBJECT COMPANY*, containing information on the targeted firm (target). To get information on these proxy communications, I begin with the filer particulars. Every institutional investment manager with at least $100 million in equity assets under management is required to file a 13F form with the SEC. Thus, I include only those filers that have filed 13F-HR, 13F-NT, or 13F-E forms to make a list of all the investment firms. Activists must file with the SEC if they discuss material information even if the information is not part of a campaign. I filter out filings that (i) do not contain text, (ii) refer to an external exhibit document, and (iii) are related to merger and acquisition, litigation, or banter (Icahn 2013). I remove duplicate filings, which are usually the same document filed by the subject company for easier access to shareholders.

I get a total of 4,159 proxy filings related to proxy fights, which include 290 DEFC14A, 3,484 DFAN14A, and 385 PREC14A filings. I combine these proxy filings if they are less than 180 days apart and have the same activist and target. However, in two cases, I combine filings that are more than 180 days apart - the 2006 Sunset Financial proxy fight and the 2018 Alpine dividend fund proxy fight. I get 533 confrontational proxy fights, over the 2004–2019 period, with an average of eight filings per proxy fight. The proxy filings contain information related to activist identification, activist's message to shareholders, voting procedure, activist's holdings in the target firm, and other legal disclosure. Sometimes the activist also discusses their portfolio, past activism success, etc. I parse out the activist's message to shareholders from each filing and combine the messages across filings to get the proxy communication. To parse out the message part, I look for cues that begin and end a message. Table 11 lists the ten most common cues.

[Table 11 about here.]

## A.4 Processing fund's information acquisition via EDGAR

The search traffic data for SEC.gov covers the period from February 2003 through June 2017. EDGAR log file data set includes information on the visitor's IP address, date, timestamp, CIK, and filing document's accession number. The IP addresses in the dataset are in version 4 (IPv4) format, which defines an IP address as a 32-bit number separated into four 8-bit numbers. A dot separates each 8-bit number, and the number between the dots could be between 0 and 255 ($2^8 - 1$). So a specific IP address, let's say BlackRock's, looks like 199.253.64.128. However, the last octet of the IP address in log files is replaced with alphabets, in a way to preserve the uniqueness of the IP address without revealing the full identity of the visitor. Thus, if Blackrock accesses the SEC.gov website from the IP address, the log file will show an entry 199.253.64.mns. In essence, the EDGAR log file dataset has a 24-bit (IP3) address for each EDGAR server activity. Fortunately, most institutions register large blocks of IP addresses. For example, BlackRock owns the IP addresses ranging from 199.242.6.0 to 199.242.6.255. As such, the IP3 addresses are often sufficient to pinpoint the registered institution.

Loughran and McDonald (2017) advise separating EDGAR requests generated by robots from server requests by regular investors. I classify an IP address as a robot if it requests more than a thousand filings in a day. I remove IP addresses classified as robots for that particular day. To include only valid EDGAR activities, I follow Drake et al. (2015) and exclude activities not related to governance research. I remove index pages (index.htm), icons (.ico), XML filings (.xml), and filings that are under 500 bytes in size. I also combine views by an IP address if they are less than five minutes apart and for the same filing.

The second part of my dataset is a lookup table from Digital Element, a geolocation data and services firm. The table contains the timestamp of IP addresses (IPv4) and registered organization names in December 2016. I use regular expressions, such as (.*blackrock.*) for BlackRock Financial Management, to get IPv4 associated with institutions. To assign IP3 blocks to institutions, I use a procedure similar to Iliev et al. (2021). If an institution owns all or a subset of the IP3 address, and no other institution owns an address from the IP3 block, I assign it to the institution. If two or more institutions own a subset of IP3 block, I assign it to the family that contains the most IP address for the IP3 block. If two institutions own an equal number of IP addresses in an IP3 block, I drop those IP3 blocks. The chances of overestimating views from assigning an entire IP3 block to an institution if they own a fraction of addresses is low, as it is unlikely for non-financial firms to access filings from SEC.gov.

Next, I look for the validity of IP3 blocks assigned to the institution. The IP address to the organization name lookup table is a snapshot from December 2016. However, institutions sometimes change their underlying technology infrastructure and, in that process, register for different IP3 blocks. To make sure that I have credible IP3 blocks, I go back quarterly from December 2016 and see what fraction of holdings do institution access through the EDGAR server. I use CRSP mutual fund data to get institution holdings. If an institution does not access more than 1% of its holdings in two consecutive quarters, I stop including the institution before the quarter. For example, Cambiar Investors accessed 1.9%, 3.3%, 0.0%, and 0.1% of its holdings in 2015Q4, 2015Q3, 2015Q2, and 2015Q1 respectively. Therefore, I exclude Cambiar Investors from my sample before June 2015. Subsequently, I match valid IP3 blocks from the organization lookup table with IP3 from EDGAR log files.

I identify proxy fight documents based on the accession number of the filing in log files and SEC's index files. To measure the number of times an institution accessed proxy fight related filings, I aggregate views for proxy fight documents during the proxy fight period, defined as the period from the first proxy communications filing to 30-day after the last proxy communications filing. The institutions' views, as measured from EDGAR log files, likely under-represent actual views. As mentioned in Bauguess et al. (2013), the EDGAR log files do not contain any requests for SEC filings from EDGAR's FTP site. Moreover, internet service providers cache frequently requested documents for future ease of reference. As such, requests for the same content that have been cached are not captured in the log file.

# B  Method

## B.1  Estimating SVR's parameter

SVR estimation requires the user to choose two hyperparameters, which control the trade-off between in-sample and out-of-sample fit: the $\epsilon$ insensitive zone and the inverse regularization parameter, c. I use an $\epsilon$ insensitive zone value of 0.001, i.e., the SVR method does not penalize the cost function if the difference between actual and predicted *Align* is less than 0.1% percent. I do not go more granular to improve computational efficiency, as differences in against management voting that are less than 0.1% does not mean much economically.

For the inverse regularization parameter, I run a horse-race amongst various values to obtain the lowest mean absolute error and mean squared error. I use a three-fold grid search algorithm for the proposal voting data to pick inverse regularization parameter, c, from $10^j$, where j ranges from -15 to +4. I focus more on c below one as the strength of the regularization is inversely proportional to c. Figure 7 shows the best performing regularization parameter for the mean squared and the mean absolute error. The regularization parameter, c = 0.0001 has the lowest mean absolute error and mean squared error for 25% and 54% of my run sample.

[Figure 7 about here.]

## B.2 The phrases that matter

SVR coefficients vary across institutions, as well as across time. Table 12 lists a few of the phrases, out of 9,832 phrases, and their coefficients for BlackRock, Fidelity, and Vanguard on December 31st, 2008, and 2018. A general theme could be seen that environmental and social issues have become more important over time. Phrases such as "climate change", "emissions", "human rights" etc. which were either negative, indicating vote in support of the management, or close to zero have become positive in 2018. The institutions are willing to vote against the management recommendations on these issues thus making the coefficient positive. BlackRock, which has been vocal about climate change, does show a higher coefficient for the phrase compared to others. "in their proxy guidelines or letter"

[Table 12 about here.]

Some of the corporate governance phrases such as "poison pill", "board declassification", which were major priorities for these institutions in the 2000s and early 2010s have become less important over time. In the last two decade, poison pills have fallen out of favor; at the end of 2019, only 25 S&P 500 public firms had an active positive pill (Eldar and Wittry 2020). Thus institutions have shifted their focus away from these topics and thus the marginal increase in support of a shareholder proposal mentioning "poison pills" has decreased. SVR also assigns very similar coefficients to phrases which are similar in meaning, as shown in Table 12 by the two board declassification phrases. As similar phrases are associated with almost identical voting behavior by the institution, the SVR assigns close coefficients to these phrases.

We also see a rise in the coefficient for some other governance phrases such as "class common stock" as tech firms such as Dropbox and Snapchat opt for dual-class shares. Classification of common stocks is an important issue for all three institutions. BlackRock submitted a petition to the New York Stock Exchange and Nasdaq to require companies to eliminate unequal voting rights enshrined in different share classes (WSJ 2018). Likewise, Fidelity and Vanguard mention in their proxy guidelines that they "generally support proposals to recapitalize multi-class share structures," and "are opposed to dual-class capitalization structures that provide disparate voting rights" (Fidelity 2019; Vanguard 2018).

## B.3 Robustness of the SVR model

### B.3.1 SVR coefficients and actual voting

At the end of the training process, SVR assigns coefficients to each of the phrases used in shareholder proposals, where a positive (negative) coefficient implies that the phrase increases (decreases) the text's alignment with institution preferences. Figure 8a plots variations in the coefficient for one such phrase - "simple majority vote" for BlackRock, Fidelity, and Vanguard across time. Based on the SVR coefficients, Fidelity is more likely to vote against management if the proposals include measures to install simple majority standards. In Figure 8b, I plot the actual against management voting for these institutions. The figure shows that Fidelity has indeed voted against management when the phrase "simple majority vote" is mentioned in the proposals.

[Figure 8 about here.]

### B.3.2 SVR coefficients follow proxy voting guidelines

Mutual funds distribute funds' prospectus to shareholders yearly, describing, among other things - risks, investment strategies, and proxy voting guidelines. Proxy guidelines across mutual funds within an institution remain mostly consistent for a given year. Therefore, to gather voting policy text for an institution, I look for the prospectus of the biggest mutual fund that is part of the institution. I search for cues such as "Proxy Voting Guidelines," "Proxy Voting Policies and Procedures," etc., to extract the proxy voting guidelines. I get 378 proxy guidelines across 48 institutions over the 2004–2018 period.

[Figure 9 about here.]

It is unclear how the presence of specific phrases in the proxy guidelines should affect phrases' SVR coefficients. Couvert (2021) finds that voting policies are a major predictor of funds' voting behavior. The occurrence of phrases such as "right to call shareholder meeting" indicates that the institution wants to implement this right and would vote against the management if shareholder proposals contain this phrase. Higher against management voting would give these phrases a more positive coefficient. For example, Figure 9 shows variations in SVR coefficients of "call special meet" for Morgan Stanley across time. When Morgan Stanley mentions more about shareholders' right to have special meetings, the corresponding SVR coefficient is also higher. On the contrary, sometimes institutions also write about issues, such as climate, environmental, social, etc., which they feel are part of management decision prerogative and thus would vote with the management on those proposals. Therefore, mentions of these phrases in proxy guidelines could mean more negative SVR coefficients for the phrases.

To circumvent the ambiguity, I look for the absolute value of the coefficients. The rationale for this choice is that when an institution mentions a particular phrase in its proxy guidelines, it is important to their voting decisions. As such, the institution would vote more consistently when those phrases occur in a proposal. The consistency in voting assigns a higher absolute value to coefficients, more positive if the phrase is about supporting shareholders and more negative if the phrase is about supporting management. In this section, I look for whether the SVR coefficients follow institution's policy guidelines by employing:

$$abs(\beta)_{n,i,t+1} = \beta Count_{n,f,t} + \delta_{i \times t} + \epsilon_{n,i,t} \tag{8}$$

where $abs(\beta)$ represents the absolute SVR coefficient associated with a phrase or ngram, $n$, for an institution, $i$, at time t + 1. *Count* is the number of times a phrase appeared in the institution's proxy guidelines text filed in year t. Since I use a two-year training period for SVR, I relate phrase counts from the proxy guidelines document to the SVR coefficients calculated at the end of next year. $\delta_{i \times t}$ shows institution cross time level fixed effect and the errors, $\epsilon_{n,i,t}$, are clustered at the institution level.

[Table 13 about here.]

Table 13 reports that SVR coefficients are in line with their mentions in proxy voting guidelines. Within an institution for a particular year, the coefficients are higher in absolute terms for phrases mentioned more in the proxy guidelines. For every mention of a phrase in the proxy guidelines, the absolute value of the coefficient increases by 0.007 percentage points. Some of the phrases that appear in the shareholder proposal may not appear in proxy guidelines of an institution. One could argue that the phrases that do not appear could indeed be less significant for voting decisions, and that is why we have a positive coefficient associated with counts. To alleviate these concerns, in Columns (4)–(6), I include only those phrases that appeared at least once in the proxy guidelines text for the institution. The results remain robust for the smaller sub-sample.

# C  Results

## C.1  Examples of activists selectively using phrases

Figure 10 reports three examples of how activists focus on issues that matter to institutions with significant voting power. I choose these examples, as they have a variation in holdings between the three big institutions. In 2009, when Ramius LLC engaged CPI Corp, the focus of the proxy fight was board members not having relevant industry experience. In the proxy communications, Ramius notes "experience board" 24 times (Ramius 2009). Incidentally, the experience of board members is important to Vanguard as well, which holds 2.8% of CPI shares.

[Figure 10 about here.]

Similarly, the 2007 Flagg Street Capital proxy fight with Pomeroy Solutions was centered on an issue important to Fidelity, which owned 11.4% of Pomeroy shares. During the two years before the proxy fight, Fidelity voted against management in 30% of shareholder proposals, compared to 25% by BlackRock and Vanguard, containing the phrase "personal benefits."

Thus, a proxy fight that discusses management's embezzlement would be closer to the preferences of Fidelity. The proxy communications discussed how the Pomeroy family had run the company for personal benefit, including the transfer of the CEO position from David Pomeroy to his son (FlaggStreet 2007). Lastly, In 2013, FrontFour Capital discussed Ferro Corporation's deteriorating operating performance and how that has reflected on stock price (FrontFour 2013). Stock price performance was an important issue for both BlackRock and Vanguard, which together owned 10.3% of Ferro Corporation shares.

## C.2 Estimates for sub-samples

[Table 14 about here.]

[Table 15 about here.]

## C.3 Alignment score is higher for experienced activists

To test whether activist's communication alignment increase with their experience, I define *NumInteraction*$_{p,f}$, which indicates the interaction count for an activist with an institution, *i*, before a particular proxy fight, *p*. I start by sorting the proxy fights in terms of proxy fight date and assigning *NumInteraction*$_{p,f}$ equal to zero. For each proxy fight by the activist, if the institution owns more than a percent of shares in the target, *NumInteraction*$_{p,f}$ increases by one. In essence, *NumInteraction* measures the number of times an activist has interacted with an institution when the institution owns significant shares in the target. I employ:

$$Align_{p,f} = \beta NumInteraction_{p,f} + \delta_p + \delta_f + \epsilon_p \tag{9}$$

where $Align_{p,i}$ is the predicted alignment of proxy communications, *p*, with the institution, *i*, preferences. *NumInteraction* is the number of times the institution has been a significant shareholder in the activist's proxy fights. $\delta_p$ and $\delta_i$ are proxy fight level and institution level fixed effects. Finally, I adjust the standard errors, $\epsilon_{p,i}$, for clustering at the proxy fight level.

Table 16 demonstrates that the activists raise issues closer to institution preferences, when they interact more with an institution. For every interaction between an activist and an institution owning more than a percent of target shares, the proxy communications' alignment with the institution is 0.9 percentage point higher. The increase is substantial, compared to the average proxy communications' alignment of 48 percentage points. The results are robust to fixing proxy fight and institution level variations. Thus, the activists are more willing to include phrases that appeal to institutions with whom they have had a proxy fight relevant interaction before.

[Table 16 about here.]

The results support Appel et al. (2019), who point out that activists have learned through their repeated interactions, and are able to tailor their campaign strategies and goals to reflect priorities of long-term investors. Howard Sherman, CEO of Institutional Shareholder Services, also agrees that "these hedge funds are looking for returns, the push for governance is coming from a larger and larger number of public pension funds and investment managers" (InstitutionalInvestor 2006). The shift in strategies could explain the increased success of activists and the increased openness of some institutions to activists' demands. For example, in the 2015 letter to corporates, Larry Fink, CEO of BlackRock, stressed that short-term thinking is getting in the way of long-term business growth. (BlackRock 2015). In contrast, Fink admitted in 2018 that the interactions between targets and activists are often productive for long-term investors like his funds (Reuters 2018).

## C.4 Limitations of text-based measure and mitigation

Certain limitations come with a text-based model. Many factors, including the content of proxy communications, firm-specific performance, general economy, relationships between institution and target, the reputation of activists, etc., play a role in how an institution votes. Failure to control for such factors could introduce an omitted variable bias that confounds inferences. Moreover, the text-based measure is likely to predict a fraction of the whole voting variation, as shown in Section 5.2. I account for these issues by using stringent fixed effects: proxy fight level and institution level.

While the voting data is available at the individual mutual fund portfolio level, I analyze voting outcomes at the proxy fight × institution level. The voting outcome at the institution level is more reasonable as an overwhelming fraction of institutions coordinate the votes across their funds Ashraf et al. (2012); Morgan et al. (2011). Thus, using an institution level outcome is more in line with the independent and identically distributed assumption on errors (Bolton et al. 2020). Moreover, instead of predicting one electoral outcome for a proxy fight, I furcate the voting at the institution level. Thus, the predicted proxy communications' alignment with institutions is correlated across a proxy fight. The correlation could reduce the standard error of $\beta$ and boost significance. Clustering at the proxy fight level mitigates this risk. I also try robust standard errors (unreported), and the results are similarly significant.

My training sample, which contains proposals at the annual meeting, is not the same as my prediction sample, the proxy communication. The discrepancy occurs because of the shortage of proxy fights that reached a voting stage. In Section 5.2, I mention that only 199 proxy fights went for voting over the 2004–2019 period, which is not enough to run a machine learning algorithm. To mitigate the differences between training and prediction sample, I filter out all the management proposal and train the SVR on shareholder proposals only. Shareholder proposals are often more in line with activist's proposals in proxy communication. Moreover, over the 2003–2018 period, institutions' voting on shareholder proposals (44% against the management) is in line with the voting in proxy fight proposals (48% against the management).

I focus on proxy fights to illustrate the importance of persuasion in institution voting. However, proxy fight proposals make a small portion of all the voting decisions. In any year, less than twenty proxy fights reach the voting stage, while prominent mutual fund institutions, on average, cast over 30,000 votes at American public companies ([NYTimes 2019](#)). Nonetheless, proxy fights have higher stakes for all parties involved compared to routine proxy votings for which investor votes are mostly precatory ([Brav et al. 2023](#); [Buchanan et al. 2012](#); [Klein and Zur 2009](#)). The proxy communications are often more informative, to the point, and without boilerplate or legal jargon. Thus, by analyzing voting in proxy fights, I provide evidence of activists successfully persuading institutions in events that have a long-term effect on the economy.

# D  Additional information for robustness tests

## D.1  Classification of proposals into types

[Table 17 about here.]

## D.2  Confidence interval for varying parameters

Results in the paper are also robust to various specification choices. In Section 5.3, I use a cutoff for mutual fund ownership to define the ownership dummy. The dummy is one, if mutual funds with voting information own more than the average ownership of 14.3%. Appendix Figure 11 shows that the results in Section 5.3 hold for different cutoff parameters. For the SVR method, I make subjective choices in terms of parameters used: (i) n-gram length = 5, (ii) threshold for excluding words with higher frequency = 0.7, (iii) minimum number of voting observations = 100, and (iv) window of shareholder proposals = 2 years. Figure 12 shows coefficients with a 95% confidence interval for Equation 3. The coefficients are significant for changing parameters on either side of the respective cutoffs. Thus, the text-based voting prediction is rooted in institutions' proxy guidelines and is insensitive to changing parameters.

[Figure 11 about here.]

[Figure 12 about here.]

**Figure 6:**
**Institutions that own significant voting power vary across proxy fights.**
Figure (a) plots institutions with the largest stock ownership in the targets at the initiation of proxy fights over the 2004–2019 period. The holdings data is gathered from CRSP, and aggregated to parent institutions that manage these funds. The total number of proxy fights is annotated at the center. Figure (b) plots the distribution of investment in stocks as a percent of market cap across proxy fights for BlackRock, Fidelity, and Vanguard.

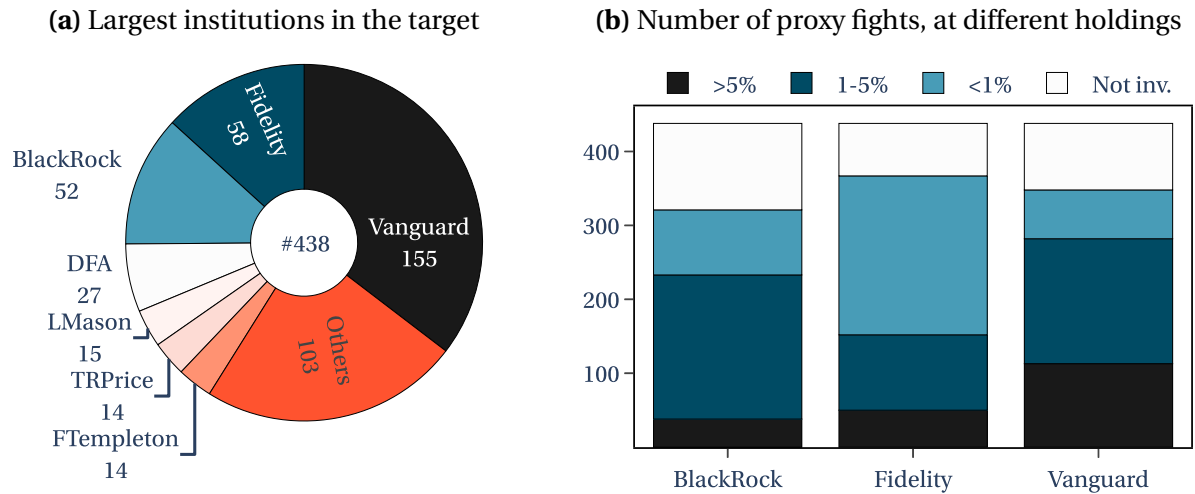

**Figure 7:**
**The SVR's inverse regularization parameter that minimizes out-of-sample errors.**
The figure plots the percent of SVR runs for which an inverse regularization parameter (or c) reduces the out-of-sample mean absolute and mean squared error. The run sample includes 25 randomly selected institutions each quarter over the 2004–2019 period, totaling to 1500 SVR runs (58 quarters * 25 institutions). I use a three-fold cross-validation via GridSearch package to find the inverse regularization parameter with the lowest error.

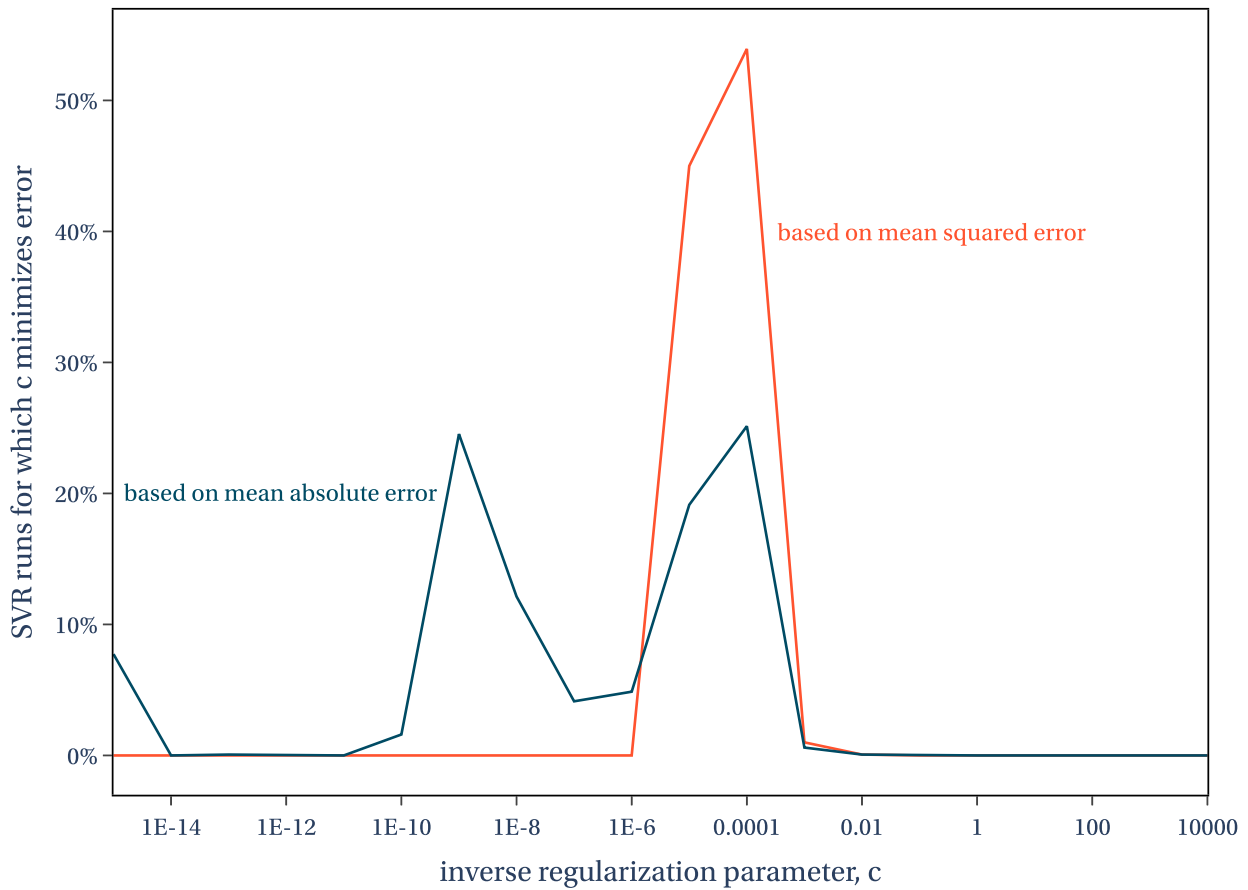

**Figure 8:**
**SVR coefficients follow proxy voting choices.**
Figure (a) plots SVR coefficients for "simple majority vote" calculated on December 31st of each year for BlackRock, Fidelity, and Vanguard. The calculations are based on the institution's proxy voting choices in shareholder proposals during the two years before the calculation date. The coefficients indicate the marginal increase in the text's alignment with institution preferences if the text contains one more instance of the phrase. For example, a coefficient of 0.004 for Fidelity in December 2010 indicates that Fidelity is 0.4 percentage points more likely to vote against the management for every instance of "simple majority vote" in the proposal text. Figure (b) plots the fraction of shareholder proposals containing "simple majority vote," where the institution voted against the management recommendation.

**(a)** SVR assigned coefficient for "simple majority vote"

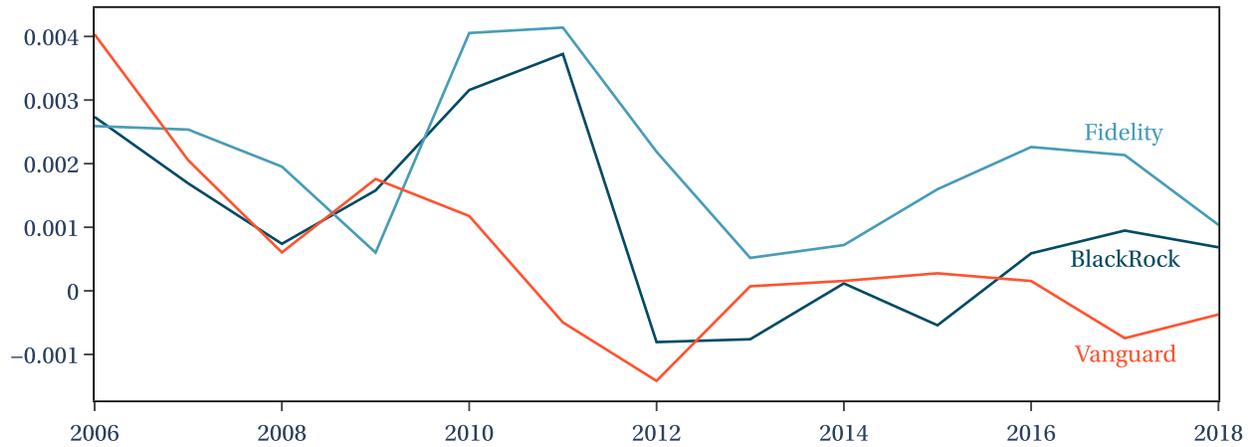

**(b)** The fraction of shareholder proposals containing "simple majority vote," where the institution voted against the management recommendation

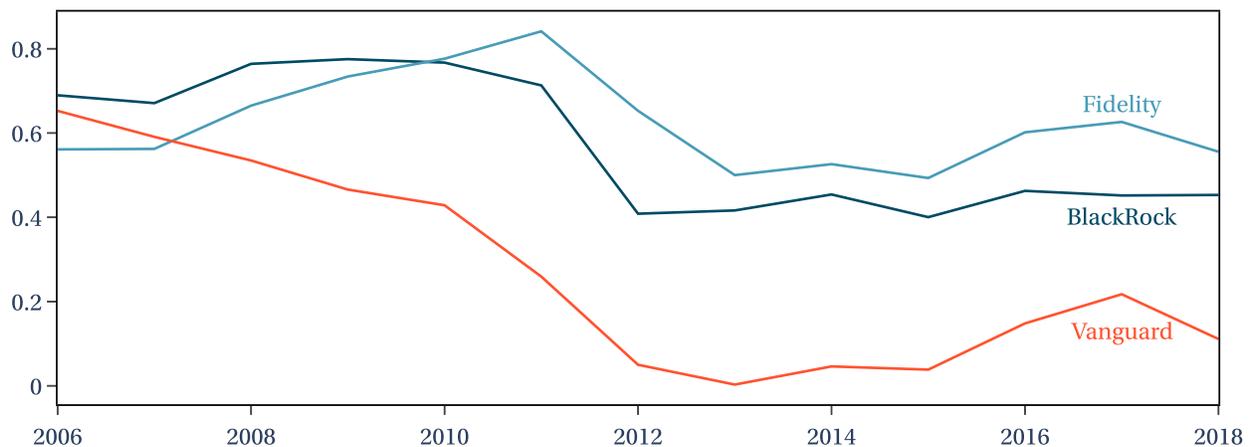

**Figure 9:**
**SVR coefficients of "call special meet" follow Morgan Stanley's proxy voting guidelines.**
This figure plots the number of times the phrase "call special meet" is used in Morgan Stanley's proxy voting guidelines and subsequent SVR coefficients. The SVR coefficients are calculated as of December 31st of the year after proxy guidelines are published. The coefficients are based on the institution's voting patterns on shareholder proposals in the two years before the calculation date.

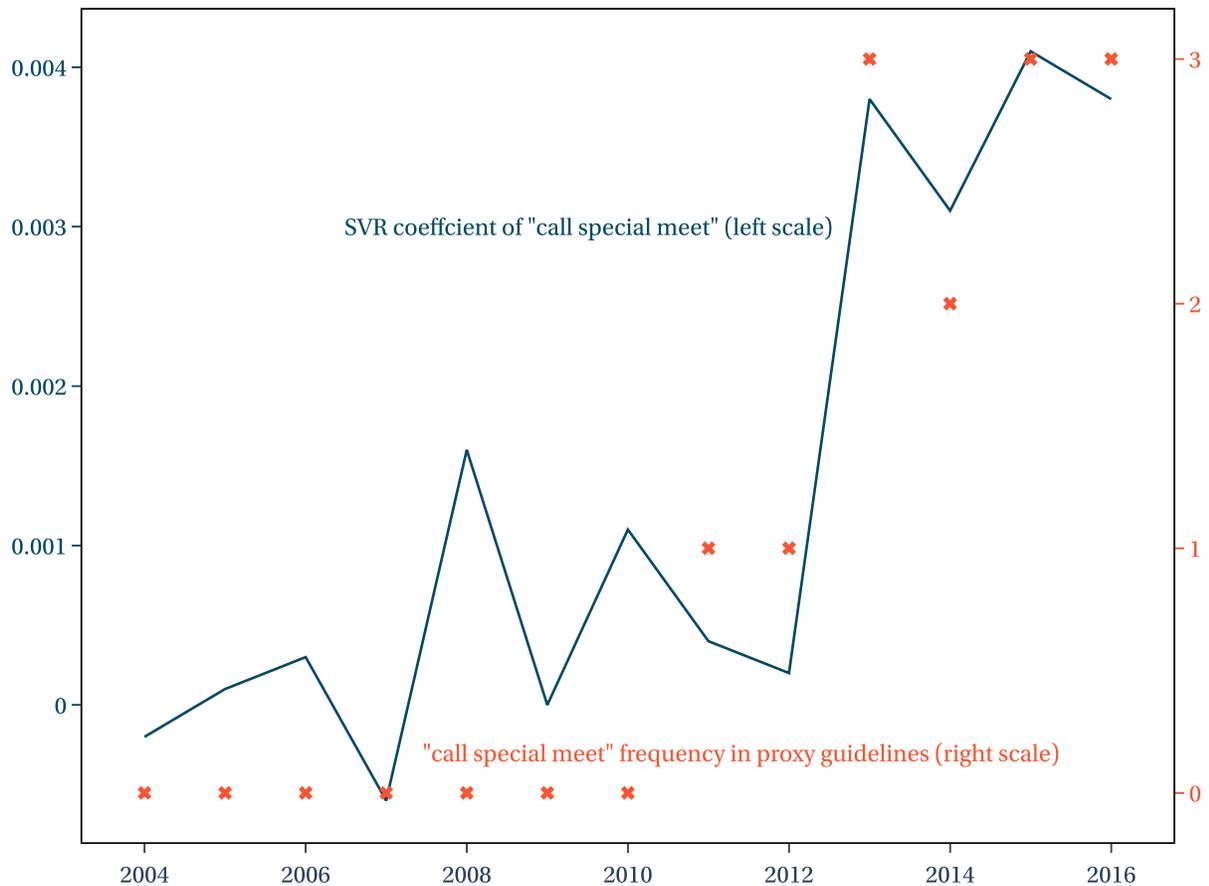

**Figure 10:**
**Activists use phrases that will increase proxy communications' alignment with larger shareholders' preferences.**
The bar chart shows the marginal increase in proxy communications' alignment with institution preferences, if the activist uses one more instance of the phrase. The measure is derived from shareholder proposals voting in the two years before each proxy fight. The vertical thin lines indicate the percent of target shares held by the institution at the start of the proxy fight. The x-axis mentions the key phrase, followed by Year Activist, Target tuple.

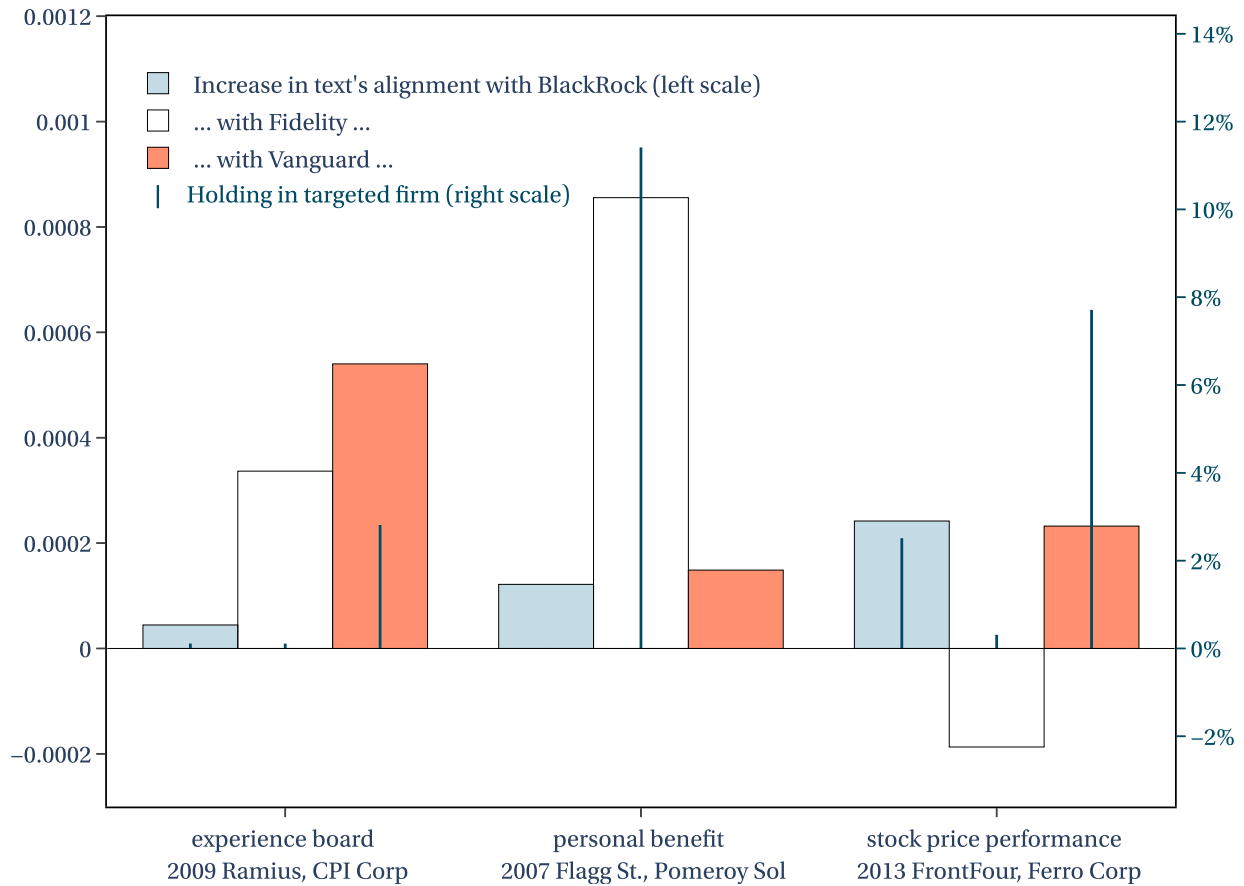

**Figure 11:**
**The positive association between proxy fight's aggregate alignment and activist's success holds for changing ownership dummy cutoff.**

This figure plots $\beta$ coefficient with 95% confidence interval for regression of proxy fight outcome on institution holdings weighted proxy communications' alignment. Specifically, I estimate:

$$Win_p = \gamma AgAlign_p + \lambda OwnDum_p + \beta AgAlign_p \times OwnDum_p + \epsilon_p$$

where $Win_p$ represents a dummy, which is one if the result of the proxy fight, $p$, is Successful or Settled. *AgAlign* is the holdings-weighted proxy communications' alignment with institutions. *OwnDum* is the ownership dummy that is one if the mutual funds, whose voting information is available, own more than the cutoff of target shares. The sample consists of proxy fights that went to a voting stage over the 2004–2019 period. The independent variable, *AgAlign*, is scaled by the standard deviation of the underlying variable. The standard errors, $\epsilon_p$, is robust and computed with the sandwich estimator of variance.

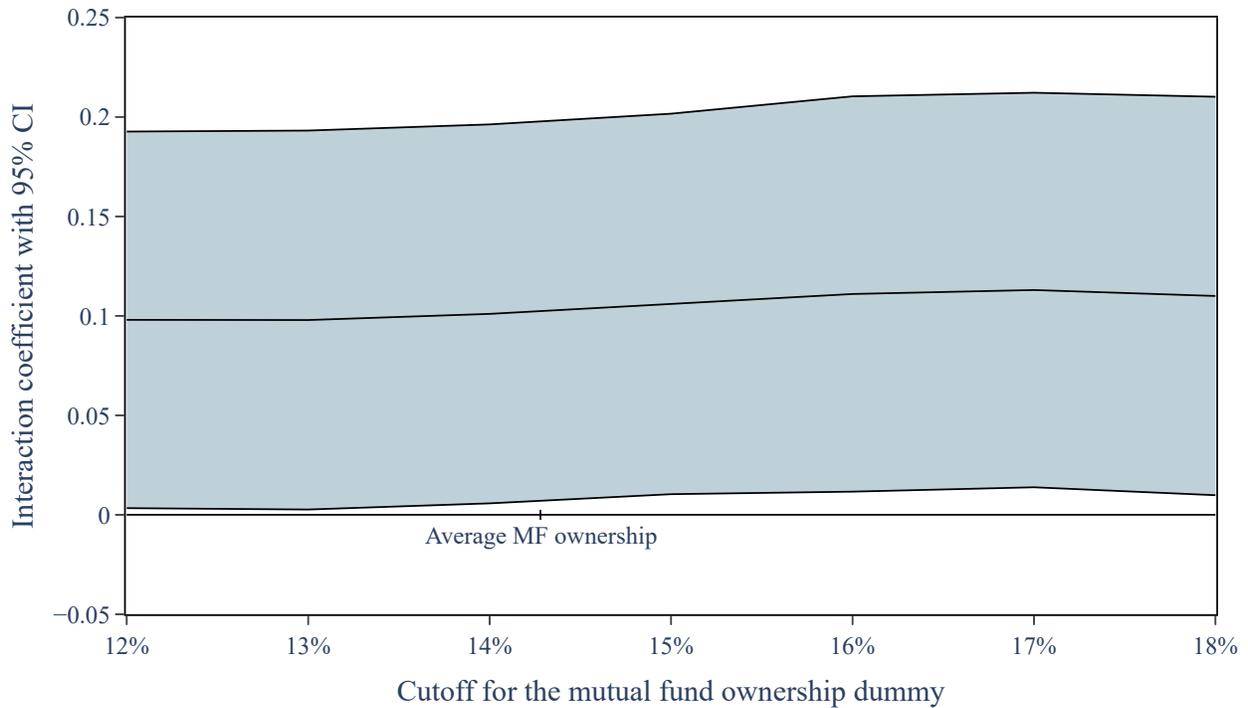

**Figure 12:**
**The positive association between institution holdings and proxy communications' alignment is robust to changing SVR parameters.**

This figure plots $\beta$ coefficient with 95% confidence interval for regression of institution's text-based likelihood of supporting activists on institution holdings in the target. Specifically, I estimate:

$$Align_{p,i} = \beta Holding_{p,i} + \delta_p + \delta_i + \epsilon_{p,i}$$

where $Align_{p,i}$ is the predicted alignment of proxy communication, $p$, with the institution, $i$, preferences. *Holding* is the percent of equity the institution owns of the target before the proxy fight, obtained from CRSP database. $\delta_p$, and $\delta_i$ represent proxy fight level and institution level fixed effects, respectively. The sample consists of proxy fights, identified using SEC filings, over the 2004–2019 period. Corresponding institutions include all the institutions that have voted in at least a hundred shareholder proposals in the two years prior to the proxy fight. The independent variable is scaled by the standard deviation of the underlying variable, meaning the coefficient can be interpreted as the effect of a one-standard-deviation change in the determinant. Standard errors, $\epsilon_{p,i}$, are clustered at the proxy fight level.

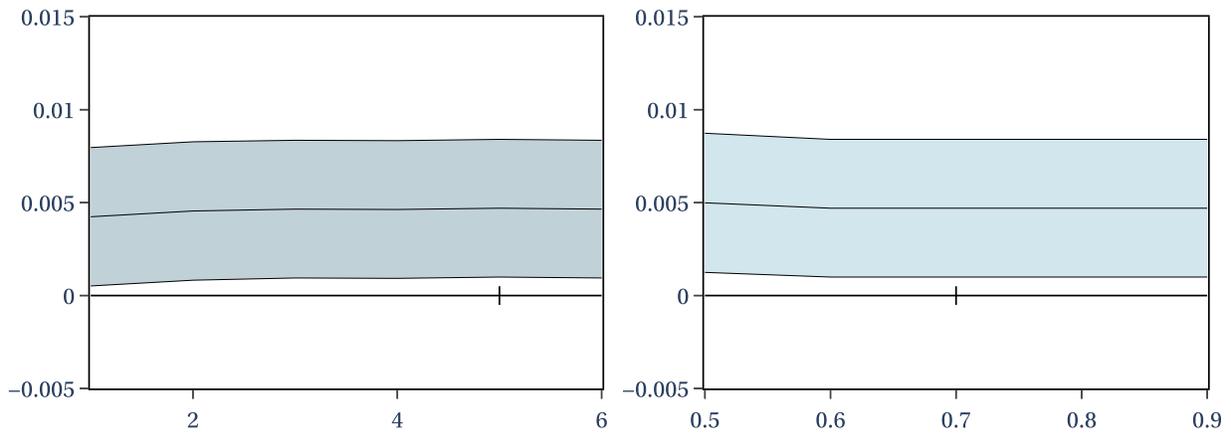

**(a)** n-gram complexity

**(b)** cuttoff for higher frequency terms

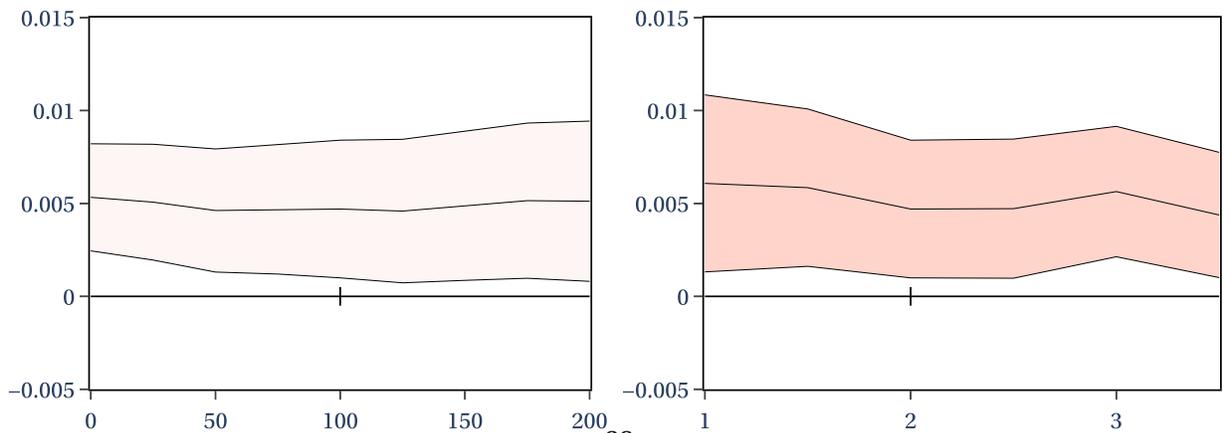

**(c)** minimum shareholder proposals required

**(d)** past voting considered (in years)

**Table 11:**
**Top ten cues to parse the message in proxy communications filings**
This table reports a subset of cues to get the message to shareholders, from a proxy communications filing. [ACTIVIST] ([TARGET]) is a placeholder for the name of activist (target), gathered from identification section in the proxy communications filing.

| Message Begin Cue | Message End Cue |
|---|---|
| Reasons for the solicitation | Sincerely yours |
| Ladies and Gentlemen | Warm regards |
| Dear Fellow Shareholder | Sincerely |
| Dear Board of Directors | Best |
| [ACTIVIST] is seeking your support for | Please sign date and return the gold proxy card today |
| The following is the text of a press release issued by [ACTIVIST] | Security holders are advised to read the proxy statement and other documents related to the solicitation of proxies |
| confirms intention to nominate [ACTIVIST] | Urge you to vote your shares on the green proxy card |
| find proxy materials for the important annual meeting of [TARGET] | Please address any correspondence to [ACTIVIST] |
| being furnished to you the stockholders of [TARGET] | For further information including full biographies of our management team |
| soliciting proxies from holders of shares of [TARGET] | Any other relevant documents are available at no charge on the secs website |

**Table 12:**
**The importance of phrases in voting decisions varies across institutions and time**
This table reports coefficients associated with phrases for three of the largest institutions in my sample. The list is based on institutions' preferences on December 31st, 2008, and 2018. The phrases are stripped of cases, punctuation, stop-words, and noun/verb forms. The coefficients are multiplied by 10,000. The coefficients indicate the marginal increase in the proxy communications' alignment with institution preferences if it contains one more instance of the phrase. For example, a coefficient of 0.008 for BlackRock indicates that BlackRock is 0.8 percentage points more likely to vote against the management for every instance of "climate change" in the proposal text.

|  | BlackRock | | Fidelity | | Vanguard | |
| --- | --- | --- | --- | --- | --- | --- |
|  | 2008 | 2018 | 2008 | 2018 | 2008 | 2018 |
| class common stock | −8.6 | 32.1 | −6.6 | 14.3 | −16.6 | 30.1 |
| climate change | −5.6 | 8.3 | −8 | −9.1 | −5.6 | −5.3 |
| declassify board | 45.1 | 9.8 | 69.6 | 7.3 | 94.7 | 8.9 |
| declassify board director | 40.8 | 3.2 | 64.3 | 3 | 76 | 3.4 |
| emission | −8 | 0 | −7 | 22.1 | 2.3 | 2.4 |
| human right | −49.3 | 2.2 | −25.1 | −1.9 | −8.6 | −6.7 |
| poison pill | 11.1 | −2.1 | 13.4 | 2.1 | 5.1 | −1.5 |

**Table 13:**
**SVR coefficients follow proxy voting guidelines**
This table reports estimates of regression of absolute phrase coefficients on the number of times the phrase appeared in the institution's proxy guidelines text. Specifically, I estimate:

$$abs(\beta)_{n,f,t+1} = \beta \, Count_{n,f,t} + \delta_{f \times t} + \epsilon_{n,f,t}$$

where $abs(\beta)$ represents the absolute SVR coefficient associated with a phrase or ngram, $n$, for an institution, $i$, at the end of year $t+1$. *Count* is the number of times the phrase appeared in the institution's proxy guidelines text filed in year t. $\delta_{f \times t}$ represents institution cross year fixed effect. (1), (2), and (3) show results for all the 9,832 phrases described in Section 3.1 for each institution. The institution sample is restricted to institutions that (i) have voted in at least a hundred shareholder proposals in the two years prior to the SVR calculation date, and (ii) have a proxy guidelines text available in 485BPOS filing. For (4), (5), and (6), I filter out phrases for an institution if the phrase is not present in any of the institution's proxy guidelines. The independent variable is scaled by the standard deviation of the underlying variable, meaning the coefficient can be interpreted as the effect of a one-standard-deviation change in the determinant. Standard errors, $\epsilon_{n,f,t}$, are clustered at the institution level, and $t$-statistics are reported in brackets below the coefficient estimates. The symbols *, **, and *** indicate significance at 10%, 5%, and 1% respectively.

|  | Absolute SVR coefficient for the phrase × 10,000 | | | | | |
| --- | --- | --- | --- | --- | --- | --- |
|  | (1) | (2) | (3) | (4) | (5) | (6) |
| Frequency of phrase | 0.736*** | 0.734*** | 0.744*** | 0.477*** | 0.476*** | 0.477*** |
| in proxy guidelines | [217.46] | [7.64] | [7.77] | [82.87] | [7.44] | [7.82] |
| Institution FE |  | Yes |  |  | Yes |  |
| Institution × year FE |  |  | Yes |  |  | Yes |
| Exclude absent phrases |  |  |  | Yes | Yes | Yes |
| Observation | 2,192,536 | 2,192,536 | 2,192,536 | 358,206 | 358,206 | 358,206 |
| $R^2$ | 0.021 | 0.04 | 0.064 | 0.019 | 0.045 | 0.075 |

**Table 14:**
**Activists tailor their communications to the preferences of large shareholders**
This table reports estimates of regression of proxy communications' alignment with the institution preferences on institutions' holdings in targets. Specifically, I estimate:

$$Align_{p,i} = \beta Holding_{p,i} + \delta_p + \delta_i + \epsilon_{p,i}$$

where $Align_{p,i}$ is the predicted alignment of proxy communication, $p$, with the institution, $i$, preferences. *Holding* is the percent of equity the institution owns of the target before the proxy fight. $\delta_p$, and $\delta_i$ represent proxy fight level and institution level fixed effects, respectively. The sample consists of proxy fights, identified using proxy communications filings, over the 2004–2019 period. Corresponding institutions include all the institutions that have voted in at least a hundred shareholder proposals in the two years prior to the proxy fight. Column (1) to (4) includes observations where an institution is invested in the target, while Column (5) to (8) excludes observations where an institution owns more than five percent of the market cap of the targeted firm. The independent variable is scaled by the standard deviation of the underlying variable, meaning the coefficient can be interpreted as the effect of a one-standard-deviation change in the determinant. Standard errors, $\epsilon_{p,i}$, are clustered at the proxy fight level, and *t*-statistics are reported in brackets below the coefficient estimates. The symbols *, **, and *** indicate significance at 10%, 5%, and 1% respectively.

|  | proxy communications' alignment with institution preferences | | | |
|---|---|---|---|---|
|  | (1) | (2) | (3) | (4) |
| Fraction of target mcap held by institution | 0.0144*** [4.05] | 0.0112*** [2.91] | 0.0125** [2.52] | 0.0065 [1.64] |
| Proxy fight FE |  | Yes |  | Yes |
| Institution FE |  |  | Yes | Yes |
| Observation | 12,579 | 12,549 | 12,560 | 12,531 |
| $R^2$ | 0.00 | 0.20 | 0.07 | 0.26 |
|  | (5) | (6) | (7) | (8) |
| Fraction of target mcap held by institution | 0.0084*** [2.97] | 0.0061* [1.81] | 0.0127*** [3.08] | 0.0085** [2.52] |
| Proxy fight FE |  | Yes |  | Yes |
| Institution FE |  |  | Yes | Yes |
| Observation | 66,154 | 66,154 | 66,154 | 66,154 |
| $R^2$ | 0.00 | 0.13 | 0.10 | 0.22 |

**Table 15:**
**Institutions conduct more research on proxy fights tailored to their preferences**
This table reports estimates of regression of institution access of proxy communications filings on the proxy communications' alignment with institution preferences. Specifically, I estimate:

$$View_{p,i} = \beta Align_{p,i} + \delta_p + \delta_i + \epsilon_{p,i}$$

where $View_{p,f}$ is the number of times an institution, $i$, accessed proxy communications filings, $p$, between the date the proxy fight begins to 30 days after the proxy fight ends. The proxy fight's beginning (end) date is based on the first (last) date of proxy fight filing by the activist. $Align_{p,i}$ is the proxy communications' alignment with institution preferences. $\delta_p$, and $\delta_i$ are proxy fight level and are institution level fixed effects. The sample consists of proxy fights, identified using proxy communications filings, and corresponding institutions over the 2004–2019 period. Data for institutions' access of filings on SEC.gov is available via DERA. Columns (4), (5), and (6) control for institution holdings in the target. Independent variables are scaled by the standard deviation of the underlying variable, meaning coefficients can be interpreted as the effects of a one-standard-deviation change in the determinant. Standard errors, $\epsilon_{p,i}$, are clustered at the proxy fight level, and $t$-statistics are reported in brackets below the coefficient estimates. The symbols *, **, and *** indicate significance at 10%, 5%, and 1% respectively.

|  | Number of times institution viewed proxy communications filings on SEC.gov | | | | | |
| --- | --- | --- | --- | --- | --- | --- |
|  | (1) | (2) | (3) | (4) | (5) | (6) |
| proxy communications' alignment | 0.266*** [4.52] | 0.298*** [3.77] | 0.306*** [4.14] | 0.246*** [4.21] | 0.274*** [3.54] | 0.302*** [4.08] |
| Fraction of target mcap held by institution |  |  |  | 0.511*** [8.73] | 0.584*** [6.43] | 0.328*** [3.7] |
| Proxy fight FE |  | Yes | Yes |  | Yes | Yes |
| Institution FE |  |  | Yes |  |  | Yes |
| Observation | 10,232 | 10,207 | 10,198 | 10,232 | 10,207 | 10,198 |
| $R^2$ | 0.002 | 0.147 | 0.226 | 0.009 | 0.156 | 0.228 |

**Table 16:**
**Activists learn from interactions with institutions**
This table reports estimates of regression of proxy communications' alignment with the institution preferences on the number of times the activist has interacted with an institution. Specifically, I estimate:

$$Align_{p,f} = \beta NumInteraction_{p,f} + \delta_p + \delta_f + \epsilon_{p,f}$$

where $Align_{p,i}$ is the predicted alignment of proxy communications, $p$, with the institution, $i$, preferences. *NumInteraction* is the number of times the institution owned more than a percent of target shares in a proxy fight initiated by the activists. The institution ownership data is obtained from the CRSP database. $\delta_p$, and $\delta_f$ are the proxy fight level and institution level fixed effects. The sample consists of proxy fights, identified using proxy communications filings, and corresponding institutions over the 2004–2019 period. Corresponding institutions include all the institutions that have voted in at least a hundred shareholder proposals in the two years prior to the proxy fight. Standard errors, $\epsilon_{p,i}$, are clustered at the proxy fight level, and $t$-statistics are reported in brackets below the coefficient estimates. The symbols *, **, and *** indicate significance at 10%, 5%, and 1% respectively.

|  | Proxy communications' alignment with institution preferences | | |
|---|---|---|---|
|  | (1) | (2) | (3) |
| Number of Interaction | 0.0098*** | 0.0084** | 0.0088** |
|  | [4.89] | [2.27] | [2.15] |
| Proxy fight FE |  | Yes | Yes |
| Institution FE |  |  | Yes |
| Observation | 66,432 | 66,432 | 66,432 |
| $R^2$ | 0 | 0.135 | 0.224 |

**Table 17:**
**Shareholder proposal classified into types**

This table classifies shareholder proposals into 25 proposal types based on their description in the ISS database. I start grouping proposals together beginning from the most frequent description; as such, the 25 types listed below cover 90% of shareholder proposals over the 2003–2018 period.

| Prop. type | General description of proposals in ISS database | # of sh. prop. |
|---|---|---|
| 1 | Elect Directors (Opposition Slate); Elect a Shareholder-Nominee to the Board (Proxy Access Nominee); Elect Director (Cumulative Voting or More Nominees Than Board Seats).; Elect a Shareholder-Nominee to the Board; Elect Director Nominated by Preferred Shareholders; Elect Directors (Bundled Dissident Slate) | 1918 |
| 2 | Require Independent Board Chairman | 663 |
| 3 | Declassify the Board of Directors | 629 |
| 4 | Political Contributions Disclosure | 530 |
| 5 | Require a Majority Vote for the Election of Directors | 498 |
| 6 | Company-Specific – Shareholder Miscellaneous | 358 |
| 7 | Amend Articles/Bylaws/Charter – Call Special Meetings | 301 |
| 8 | Advisory Vote to Ratify Named Executive Officers' Compensation | 289 |

**Table 17 – continued from previous page**

| Prop. type | General description of proposals in ISS database | # of sh. prop. |
|---|---|---|
| 9 | Company Specific-Governance Related; Company-Specific Board-Related; Amend Articles/Bylaws/Charter – Non-Routine.; Approve Recapitalization Plan for all Stock to Have One-vote per Share; Eliminate or Restrict Severance Agreements (Change-in-Control); Amend Vote Requirements to Amend Articles/Bylaws/Charter; Establish Other Governance Board Committee; Adopt Proxy Access Right; Establish Environmental/Social Issue Board Committee; Require Director Nominee Qualifications (Excluding Environmental & Social); Submit SERP to Shareholder Vote; Change Size of Board of Directors; Establish Term Limits for Directors; Proxy Voting Tabulation; Amend Proxy Access Right; Approve/Amend Terms of Existing Poison Pill; Require More Director Nominations Than Open Seats; Require Majority of Independent Directors on Board; Amend Articles/Bylaws/Charter to Remove Antitakeover Provisions; Elect Supervisory Board Members (Bundled).; Elect a Shareholder-Nominee to the Supervisory Board.; Proxy Voting Disclosure, Confidentiality, and Tabulation; Amend articles/bylaws/charter – Filling Vacancies; Require Environmental/Social Issue Qualifications for Director Nominees; Adopt Policy on Succession Planning; Amend Articles Board-Related; Establish SERP Policy; Reimburse Proxy Contest Expenses; Limit Composition of Committee(s) to Independent Directors; Establish Director Stock Ownership Requirement; Provide for Confidential Voting (INACTIVE); ... | 1184 |
| 9 | Amend Articles/Bylaws/Charter – Removal of Directors; Amend Articles/Charter Equity-Related.; Eliminate or Restrict Shareholder Rights Plan (Poison Pill); Establish a Compensation Committee; Rotate Annual Meeting Location; Establish Shareholder Advisory Committee; Proxy Voting Disclosure; Establish Mandatory Retirement Age for Directors; Establish a Nominating Committee; Restore Preemptive Rights of Shareholders (INACTIVE) | 1184 |
| 10 | Restore or Provide for Cumulative Voting | 245 |
| 11 | Proxy Access | 232 |
| 12 | Submit Shareholder Rights Plan (Poison Pill) to Shareholder Vote | 210 |

| Prop. type | General description of proposals in ISS database | # of sh. prop. |
|---|---|---|
| 13 | Stock Retention/Holding Period; Double Trigger on Equity Plans; Compensation- Miscellaneous Company Specific; Limit/Prohibit Executive Stock-Based Awards; Review Executive Compensation (INACTIVE); Pay For Superior Performance; Report on Pay Disparity; Clawback of Incentive Payments; Miscellaneous – Equity Related; Expense Stock Options (INACTIVE); Limit Executive Compensation; Link Executive Pay to Social Criteria; Increase Disclosure of Executive Compensation; Death Benefits/Golden Coffins; Non-Employee Director Compensation; Disclose Information on Compensation Consultant; Put Repricing of Stock Options to Shareholder Vote; Adjust Executive Compensation Metrics for Share Buybacks | 1214 |
| 14 | Remove Existing Directors | 202 |
| 15 | Political Lobbying Disclosure | 200 |
| 16 | Provide Right to Act by Written Consent | 199 |
| 17 | Social Proposal | 196 |
| 18 | Improve Human Rights Standards or Policies | 189 |
| 19 | Performance-Based and/or Time-Based Equity Awards | 188 |
| 20 | Reduce Supermajority Vote Requirement | 173 |
| 21 | Report on Sustainability; GHG Emissions; Climate Change; Community-Environmental Impact; Report on Climate Change; Animal Welfare; Renewable Energy; Report on Environmental Policies; Recycling; Nuclear Power - Related; Environmental - Related Miscellaneous (INACTIVE); Energy Efficiency; Toxic Emissions; Toxic Substances (INACTIVE) | 793 |
| 22 | Appoint Alternate Internal Statutory Auditor(s) [and Approve Auditor's/Auditors'Remuneration].; Limit Auditor from Providing Non-Audit Services; Auditor Rotation; Appoint Internal Statutory Auditor(s) Nominated by Preferred Shareholders [and Approve Auditor's/Auditors' Remuneration] | 85 |
| 23 | Adopt Sexual Orientation Anti-bias Policy | 106 |
| 24 | Submit Severance Agreement (Change-in-Control) to Shareholder Vote | 92 |
| 25 | Board Diversity; Report on EEO | 125 |